\documentclass[a4paper,11pt]{article}
\usepackage[pdftex]{graphicx,xcolor}
\usepackage{latexsym}
\usepackage{amsmath,amssymb,amsfonts,amsthm}
\usepackage{authblk}
\usepackage{hyperref}
\usepackage{subcaption}
\usepackage{empheq}
\usepackage{float}
\usepackage{xcolor}
\hypersetup{
    colorlinks=true,
    citecolor=blue,
    linkcolor=red,
    urlcolor=green,
}
\usepackage{algorithm,algpseudocode}
\usepackage{url}
\usepackage[utf8]{inputenc}
\usepackage{tikz}
\usetikzlibrary{positioning}
\usetikzlibrary{arrows.meta}
\usetikzlibrary{calc,shapes.callouts,shapes.arrows,patterns}
\usepackage{pxpgfmark}
\definecolor{blk}{RGB}{63,63,63}
\definecolor{wht}{RGB}{255,255,255}
\title{
Computational Complexity and Integer Programming Formulation of the Oredango Puzzle
}
\author[1]{Takuma Takahata}
\author[1]{Norito Minamikawa}
\author[1]{Takayuki Okuno}
\affil[1]{Faculty of Science and Engineering, Seikei University}
\affil[ ]{\texttt {\href{mailto:norito-minamikawa@st.seikei.ac.jp}{norito-minamikawa@st.seikei.ac.jp}}}
\theoremstyle{plain}
\newtheorem{thm}{Theorem}[section]

\newtheorem{example}[thm]{Example}

\newcommand{\inc}{{\rm inc}}
\newcommand{\norm}[1]{\lVert#1\rVert}
\newcommand{\supp}[1]{ {\rm supp} ( #1 ) }
\newcommand{\dom}{{\rm dom}}
\newcommand{\suppp}[1]{ {\rm supp}^{+} ( #1 ) }
\newcommand{\suppm}[1]{ {\rm supp}^{-} ( #1 ) }
\newcommand{\univec}{U}
\def\AlgorithmText#1{\rm \textbf{Algorithm} #1} 
\begin{document}

\maketitle

\vspace{-7mm}
%------------------------ Abstract ------------------------%
\begin{abstract}
Oredango puzzle, one of the pencil puzzles, 
was originally created by Kanaiboshi and published in the popular puzzle magazine Nikoli.
In this paper, we show NP- and ASP-completeness of Oredango 
by constructing a reduction from the 1-in-3SAT problem. 
Next, we formulate Oredango as an 0-1 integer-programming problem, 
and present numerical results obtained by solving Oredango puzzles from Nikoli and PuzzleSquare JP using a 0-1 optimization solver.
\end{abstract}

\section{Introduction} \label{sec:intro}
The pencil puzzle is a type of puzzle game where solutions are filled in with a pencil according to a given rule. In this paper, we discuss the {\it  Oredango} puzzle, which was originally created by Kanaiboshi and published in the popular puzzle magazine {\it Nikoli} (Vol. 184, No. 9, 2023). 
Our aim is two-fold: 
first, to evaluate the computational complexity of solving Oredango;
second, to establish its 0-1 integer programming approach to solve Oredango, 
formulate all the Oredango puzzles from Nikoli 
as 0-1 integer programs and examine the elapsed time for solving them with an optimization solver.  

In Oredango, we are given a rectangular grid of size $m \times n$ as input, where white circles are placed in some cells, along with broken lines called {\it skewers} connecting circles.
For convenience, a circle that is not connected to any other circles
is regarded as being in a skewer of length zero.
A skewer of nonzero length can directly connect only two adjacent circles.
Nonnegative integers are written inside some circles. Each circle contains at most one integer, and each skewer contains at most one circle with an integer. 

The goal of Oredango is to color white circles black so that the following conditions are fulfilled: 
\begin{enumerate} 
\item[(a)] In any skewer, the number of black-colored circles equals the integer inside the circles connected by that skewer. 
\item[(b)] In any skewer, there are no three consecutive circles of the same color, that is, white or black. 
\item[(c)] In any row of the grid, the same condition as (b) holds. 
\item[(d)] In any column of the grid, the same condition as (b) and (c) holds.
\end{enumerate}

\noindent 
In addition, Oredango allows for cells without circles. If empty cells exist, rules (c) and (d) are applied while skipping such cells. Figure~\ref{fig:ex} illustrates an Oredango puzzle on the left side, along with the correct answer on the right side, while Figure~\ref{fig:ex2} presents two incorrect answers to the same puzzle. In the left of Figure~\ref{fig:ex2}, rules (c) and (d) are satisfied, but neither (a) nor (b) because only three black circles are in the skewer containing the number 4 and three consecutive black circles are in the skewer containing the number 3. Moreover, the right figure of Figure~\ref{fig:ex2} satisfies (a) and (b), but both (c) and (d) are violated, as circles with the same colors appear in three consecutive columns in the third and fourth rows, respectively, and black circles appear in three consecutive rows in the third column, 
wherein the empty cell in the third row is skipped. 

\begin{figure}[t]
\centering
\begin{tikzpicture}[scale=0.8]
%grid
\draw[step=1.0cm,color={rgb:black,1;white,4}] (0,0) grid (4.0,4.0);
%frame
\draw[line width=0.6mm] (0,0) -- (0,4.0);
\draw[line width=0.6mm] (4.0,0) -- (4.0,4.0);
\draw[line width=0.6mm] (0,0) -- (4.0,0);
\draw[line width=0.6mm] (0,4.0) -- (4.0,4.0);
%skewers
\draw[line width=0.6mm] (0.5,0.5) -- (1.5,1.5);
\draw[line width=0.6mm] (1.5,1.5) -- (0.5,2.5);
\draw[line width=0.6mm] (0.5,2.5) -- (1.5,3.5);
\draw[line width=0.6mm] (1.5,3.5) -- (2.5,3.5);
\draw[line width=0.6mm] (0.5,1.5) -- (1.5,0.5);
\draw[line width=0.6mm] (1.5,0.5) -- (2.5,0.5);
\draw[line width=0.6mm] (2.5,0.5) -- (3.5,1.5);
\draw[line width=0.6mm] (3.5,1.5) -- (2.5,2.5);
\draw[line width=0.6mm] (2.5,2.5) -- (3.5,3.5);
%circles
\node[draw,circle,minimum size=0.6cm, fill=wht] at (0.5,0.5) {};
\node[draw,circle,minimum size=0.6cm, fill=wht] at (1.5,0.5) {};
\node[draw,circle,minimum size=0.6cm, fill=wht] at (2.5,0.5) {};
\node[draw,circle,minimum size=0.6cm, fill=wht] at (3.5,0.5) {};
\node[draw,circle,minimum size=0.6cm, fill=wht] at (0.5,1.5) {};
\node[draw,circle,minimum size=0.6cm, fill=wht] at (1.5,1.5) {};
\node[draw,circle,minimum size=0.6cm, fill=wht] at (3.5,1.5) {};
\node[draw,circle,minimum size=0.6cm, fill=wht] at (0.5,2.5) {};
\node[draw,circle,minimum size=0.6cm, fill=wht] at (2.5,2.5) {};
\node[draw,circle,minimum size=0.6cm, fill=wht] at (0.5,3.5) {};
\node[draw,circle,minimum size=0.6cm, fill=wht] at (1.5,3.5) {};
\node[draw,circle,minimum size=0.6cm, fill=wht] at (2.5,3.5) {};
\node[draw,circle,minimum size=0.6cm, fill=wht] at (3.5,3.5) {};
%numbers
\node[circle,black] at (0.5,2.5) {\Large 3};
\node[circle,black] at (3.5,3.5) {\Large 4};
\node[circle,black] at (3.5,0.5) {\Large 1};
\node[circle,black] at (0.5,3.5) {\Large 0};
\draw [line width=0.5mm, arrows = {-Stealth[scale=1.5]}]  (4.5,2.0) -- (5.5,2.0);
\end{tikzpicture}
\hspace{0.1cm}
\begin{tikzpicture}[scale=0.8]
%grid
\draw[step=1.0cm,color={rgb:black,1;white,4}] (0,0) grid (4.0,4.0);
%frame
\draw[line width=0.6mm] (0,0) -- (0,4.0);
\draw[line width=0.6mm] (4.0,0) -- (4.0,4.0);
\draw[line width=0.6mm] (0,0) -- (4.0,0);
\draw[line width=0.6mm] (0,4.0) -- (4.0,4.0);
%skewers
\draw[line width=0.6mm] (0.5,0.5) -- (1.5,1.5);
\draw[line width=0.6mm] (1.5,1.5) -- (0.5,2.5);
\draw[line width=0.6mm] (0.5,2.5) -- (1.5,3.5);
\draw[line width=0.6mm] (1.5,3.5) -- (2.5,3.5);
\draw[line width=0.6mm] (0.5,1.5) -- (1.5,0.5);
\draw[line width=0.6mm] (1.5,0.5) -- (2.5,0.5);
\draw[line width=0.6mm] (2.5,0.5) -- (3.5,1.5);
\draw[line width=0.6mm] (3.5,1.5) -- (2.5,2.5);
\draw[line width=0.6mm] (2.5,2.5) -- (3.5,3.5);
%circles
\node[draw,circle,minimum size=0.6cm, fill=blk] at (0.5,0.5) {};
\node[draw,circle,minimum size=0.6cm, fill=wht] at (1.5,0.5) {};
\node[draw,circle,minimum size=0.6cm, fill=blk] at (2.5,0.5) {};
\node[draw,circle,minimum size=0.6cm, fill=blk] at (3.5,0.5) {};
\node[draw,circle,minimum size=0.6cm, fill=blk] at (0.5,1.5) {};
\node[draw,circle,minimum size=0.6cm, fill=blk] at (1.5,1.5) {};
\node[draw,circle,minimum size=0.6cm, fill=wht] at (3.5,1.5) {};
\node[draw,circle,minimum size=0.6cm, fill=wht] at (0.5,2.5) {};
\node[draw,circle,minimum size=0.6cm, fill=blk] at (2.5,2.5) {};
\node[draw,circle,minimum size=0.6cm, fill=wht] at (0.5,3.5) {};
\node[draw,circle,minimum size=0.6cm, fill=blk] at (1.5,3.5) {};
\node[draw,circle,minimum size=0.6cm, fill=wht] at (2.5,3.5) {};
\node[draw,circle,minimum size=0.6cm, fill=blk] at (3.5,3.5) {};
%numbers
\node[circle,blk] at (0.5,2.5) {\Large 3};
\node[circle,wht] at (3.5,3.5) {\Large 4};
\node[circle,wht] at (3.5,0.5) {\Large 1};
\node[circle,blk] at (0.5,3.5) {\Large 0};
\end{tikzpicture}
\caption{An example of a $4 \times 4$ Oredango (left) and its solution (right)}
\label{fig:ex}
\vspace{-2mm}
\end{figure}
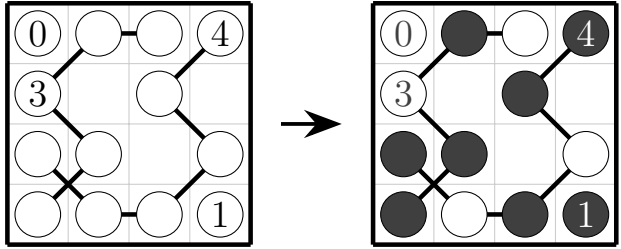
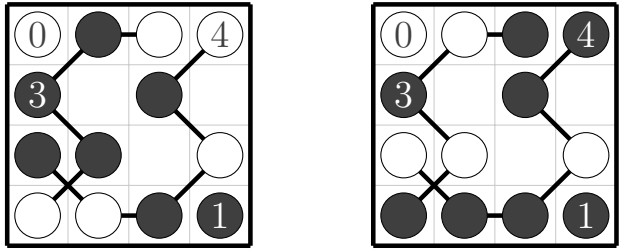
\begin{figure}[t]
\centering
\begin{tikzpicture}[scale=0.8]
%grid
\draw[step=1.0cm,color={rgb:black,1;white,4}] (0,0) grid (4.0,4.0);
%frame
\draw[line width=0.6mm] (0,0) -- (0,4.0);
\draw[line width=0.6mm] (4.0,0) -- (4.0,4.0);
\draw[line width=0.6mm] (0,0) -- (4.0,0);
\draw[line width=0.6mm] (0,4.0) -- (4.0,4.0);
%skewers
\draw[line width=0.6mm] (0.5,0.5) -- (1.5,1.5);
\draw[line width=0.6mm] (1.5,1.5) -- (0.5,2.5);
\draw[line width=0.6mm] (0.5,2.5) -- (1.5,3.5);
\draw[line width=0.6mm] (1.5,3.5) -- (2.5,3.5);
\draw[line width=0.6mm] (0.5,1.5) -- (1.5,0.5);
\draw[line width=0.6mm] (1.5,0.5) -- (2.5,0.5);
\draw[line width=0.6mm] (2.5,0.5) -- (3.5,1.5);
\draw[line width=0.6mm] (3.5,1.5) -- (2.5,2.5);
\draw[line width=0.6mm] (2.5,2.5) -- (3.5,3.5);
%circles
\node[draw,circle,minimum size=0.6cm, fill=wht] at (0.5,0.5) {};
\node[draw,circle,minimum size=0.6cm, fill=wht] at (1.5,0.5) {};
\node[draw,circle,minimum size=0.6cm, fill=blk] at (2.5,0.5) {};
\node[draw,circle,minimum size=0.6cm, fill=blk] at (3.5,0.5) {};
\node[draw,circle,minimum size=0.6cm, fill=blk] at (0.5,1.5) {};
\node[draw,circle,minimum size=0.6cm, fill=blk] at (1.5,1.5) {};
\node[draw,circle,minimum size=0.6cm, fill=wht] at (3.5,1.5) {};
\node[draw,circle,minimum size=0.6cm, fill=blk] at (0.5,2.5) {};
\node[draw,circle,minimum size=0.6cm, fill=blk] at (2.5,2.5) {};
\node[draw,circle,minimum size=0.6cm, fill=wht] at (0.5,3.5) {};
\node[draw,circle,minimum size=0.6cm, fill=blk] at (1.5,3.5) {};
\node[draw,circle,minimum size=0.6cm, fill=wht] at (2.5,3.5) {};
\node[draw,circle,minimum size=0.6cm, fill=wht] at (3.5,3.5) {};
%numbers
\node[circle,wht] at (0.5,2.5) {\Large 3};
\node[circle,blk] at (3.5,3.5) {\Large 4};
\node[circle,wht] at (3.5,0.5) {\Large 1};
\node[circle,blk] at (0.5,3.5) {\Large 0};
\end{tikzpicture}
\hspace{1.3cm}
\begin{tikzpicture}[scale=0.8]
%grid
\draw[step=1.0cm,color={rgb:black,1;white,4}] (0,0) grid (4.0,4.0);
%frame
\draw[line width=0.6mm] (0,0) -- (0,4.0);
\draw[line width=0.6mm] (4.0,0) -- (4.0,4.0);
\draw[line width=0.6mm] (0,0) -- (4.0,0);
\draw[line width=0.6mm] (0,4.0) -- (4.0,4.0);
%skewers
\draw[line width=0.6mm] (0.5,0.5) -- (1.5,1.5);
\draw[line width=0.6mm] (1.5,1.5) -- (0.5,2.5);
\draw[line width=0.6mm] (0.5,2.5) -- (1.5,3.5);
\draw[line width=0.6mm] (1.5,3.5) -- (2.5,3.5);
\draw[line width=0.6mm] (0.5,1.5) -- (1.5,0.5);
\draw[line width=0.6mm] (1.5,0.5) -- (2.5,0.5);
\draw[line width=0.6mm] (2.5,0.5) -- (3.5,1.5);
\draw[line width=0.6mm] (3.5,1.5) -- (2.5,2.5);
\draw[line width=0.6mm] (2.5,2.5) -- (3.5,3.5);
%circles
\node[draw,circle,minimum size=0.6cm, fill=blk] at (0.5,0.5) {};
\node[draw,circle,minimum size=0.6cm, fill=blk] at (1.5,0.5) {};
\node[draw,circle,minimum size=0.6cm, fill=blk] at (2.5,0.5) {};
\node[draw,circle,minimum size=0.6cm, fill=blk] at (3.5,0.5) {};
\node[draw,circle,minimum size=0.6cm, fill=wht] at (0.5,1.5) {};
\node[draw,circle,minimum size=0.6cm, fill=wht] at (1.5,1.5) {};
\node[draw,circle,minimum size=0.6cm, fill=wht] at (3.5,1.5) {};
\node[draw,circle,minimum size=0.6cm, fill=blk] at (0.5,2.5) {};
\node[draw,circle,minimum size=0.6cm, fill=blk] at (2.5,2.5) {};
\node[draw,circle,minimum size=0.6cm, fill=wht] at (0.5,3.5) {};
\node[draw,circle,minimum size=0.6cm, fill=wht] at (1.5,3.5) {};
\node[draw,circle,minimum size=0.6cm, fill=blk] at (2.5,3.5) {};
\node[draw,circle,minimum size=0.6cm, fill=blk] at (3.5,3.5) {};
%numbers
\node[circle,wht] at (0.5,2.5) {\Large 3};
\node[circle,wht] at (3.5,3.5) {\Large 4};
\node[circle,wht] at (3.5,0.5) {\Large 1};
\node[circle,blk] at (0.5,3.5) {\Large 0};
\end{tikzpicture}
\caption{Examples of wrong answers to the input of Figure~\ref{fig:ex}}
\label{fig:ex2}
\vspace{-4mm}
\end{figure}

There have been many computational complexity studies on popular games and puzzles. 
For example, Hearn and Demaine \cite{Hearn2009} surveyed related research up until the 2000s, 
while Uehara \cite{Uehara2023} and the personal webpage of Ruangwises \cite{RuangwisesHP} 
cover various pencil puzzles from the 2010s to the present. 
Sudoku, one of the most renowned pencil puzzles, 
was shown to be NP-complete and also ASP-complete \cite{Yato2003}. 
The precise definition of the ASP-complete will be explained later. 
Nondango, another pencil puzzle played by coloring circles on a rectangular grid like Oredango, 
was also proven to be NP-complete \cite{Ruangwises2024}.

To the best of the authors' knowledge, there has been no research on the computational complexity of Oredango. In this paper, we will elucidate this property. To this end, we start by considering the following decision problem from Oredango:
\begin{quote}
Oredango Decision Problem\\
\textbf{Instance:} 
Circles, skewers placed on $m \times n$ grid,
and integers within some circles \\
\textbf{Question:} Is there a way of coloring circles that satisfies rules\,(a)-(d)?
\end{quote}
In order to analyze the complexity of solving Oredango, it is sufficient to analyze the complexity of the decision problem described above. However, from the perspective of a puzzle creator, we must also consider the possibility that other solutions exist beyond the ones we have already obtained. This leads to the following problem of deciding whether another solution exists:
\begin{quote}
Oredango $n$-another decision problem\\
\textbf{Instance:}
Inputs of Oredango and $n$ solutions of Oredango satisfying rules (a)-(d)\\
\textbf{Question:}
Does there exist another solution beyond the input $n$ solutions? 
\end{quote}

\noindent
When $n=0$, this $n$-another decision problem is conventionally regarded as equivalent to the Oredango Decision Problem.  
We say that Oredango is ASP-complete if the $n$-another decision problem is NP-complete for any $n\ge 0$.
In this paper, 
we prove that Oredango is indeed ASP-complete 
not only for general inputs but also for restricted cases where
the length of each skewer and the integers in the circles
are at most 1.
Formally, we will prove the following theorem:
\begin{thm}
\label{th:main}
Oredango is ASP-complete. 
Moreover, it remains ASP-complete even when restricted to instances
where the length of each skewer and the integer in each circle are one or zero.
Here, if the length of a skewer is zero, it means there exists no skewer.
\end{thm}

Next, we formulate Oredango as a 0-1 integer programming problem and 
actually solve it using an optimization solver.
Although 0-1 integer programming is one of Karp's 21 NP-complete problems, 
numerous high-performance solvers have been developed. 
According to some numerical studies \cite{Demaine2014,Ishihama2013}, 
certain puzzles can be solved very efficiently
by using such solvers, especially when combined with appropriately modified formulations.
In this paper, using the Gurobi Optimizer \cite{Gurobi} as a 0-1 optimization solver, 
we solve a total of 36 Oredango puzzles from Nikoli \cite{Nikoli}
and PuzzleSquare JP \cite{PS}. The obtained results show that all the problems can be solved within one second. 

\section{Proof of Theorem \ref{th:main}}\label{sec:proof}
In this section, we give the proof of Theorem~\ref{th:main}.
Our proof is by a reduction from the 1-in-3SAT problem (1-in-3SAT for short), 
which is known to be one of the ASP-complete problems~\cite{Yato2003}.
In order to prove Theorem~\ref{th:main}, 
it is enough to show the latter claim, since 
the former one is readily obtained from the latter one. 
For this purpose, we will construct the reduction from 
1-in-3SAT to Oredango with the following properties carefully:
\begin{quote}
There is a one-to-one correspondence between the solution sets of 
the reduced Oredango and the original 1-in-3SAT instance.
The length of each skewer is either one or zero, and the same applies to the integer inside each circle if it exists.
\end{quote}

Let us explain the 1-in-3SAT. 
Let $x_1,x_2,\ldots,x_n$ be boolean variables which take either 0 (false) or 1 (true),
and let $U=\{x_1, x_2, \ldots, x_n\}$.
For a variable $x$, we define $\overline{x}$ as the negation of $x$,
that is, $\overline{x}=0$ (resp. 1) if $x=1$ (resp. 0). 
If $x \in U$, $x$ and $\overline{x}$ are called {\it literals} of $U$,
and a set of literals for $U$ is called a {\it clause} for $U$.
We refer to the choice of the three literals in a clause as the {\it constraint} of the clause.  
A {\it truth assignment} for $U$ stands for setting 
0 or 1 to each variable in $U$. 
Given a truth assignment for $U$,  we say that a clause is satisfied
with the truth assignment when exactly one of the literals in the clause is 1 (true).

Let $C_1,C_2,\ldots,C_m$ be clauses for $U$ such that
$|C_i|=3$ for each $i$. 
Given a collection $\mathcal{C} = \{C_1, C_2, \ldots, C_m\}$ as an instance,
the 1-in-3SAT is to
determine whether there exists some truth assignment for $U$ that 
satisfies all the clauses $C_1, C_2, \ldots, C_m$ simultaneously.

\begin{example}
\label{ex:1in3sat}
Let $U = \{x_1, x_2, x_3, x_4\}$ and $\mathcal{C} = \{C_1, C_2, C_3\}$ where 
\[
C_1 = \{x_1, x_2, x_3\}, \ C_2 = \{\overline{x_1}, x_3, x_4\}, \  C_3 = \{x_2, \overline{x_3}, \overline{x_4}\}.
\]
The following truth assignment is one of the solutions to this 1-in-3SAT:
\[
x_1 = 1, \quad
x_2 = 0, \quad
x_3 = 0, \quad
x_4 = 1.
\]
\end{example}
\subsection{Idea of the Proof} \label{subsec:idea}
We present an idea for constructing a polynomial-time one-to-one reduction 
from the 1-in-3SAT to Oredango.
We use the problem of Example~\ref{ex:1in3sat} for the sake of illustration, 
but note that the construction manner can be extended to general case immediately.

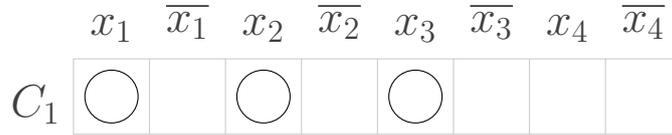
\begin{figure}[hb]
\centering
\begin{tikzpicture}[scale=1.0]
%grid
\draw[step=1.0cm, color={rgb:black, 1;white, 4}] (0.0, 0) grid (8.0, 1.0);
%circles
\node[draw, circle, minimum size=0.7cm, fill=wht] at (0.5, 0.5) {};
\node[draw, circle, minimum size=0.7cm, fill=wht] at (2.5, 0.5) {};
\node[draw, circle, minimum size=0.7cm, fill=wht] at (4.5, 0.5) {};
%variables, clauses
\node[circle, blk] at (-0.5, 0.4) {\LARGE $C_1$};
\node[circle, blk] at (0.5, 1.4) {\LARGE $x_1$};
\node[circle, blk] at (1.5, 1.5) {\LARGE $\overline{x_1}$};
\node[circle, blk] at (2.5, 1.4) {\LARGE $x_2$};
\node[circle, blk] at (3.5, 1.5) {\LARGE $\overline{x_2}$};
\node[circle, blk] at (4.5, 1.4) {\LARGE $x_3$};
\node[circle, blk] at (5.5, 1.5) {\LARGE $\overline{x_3}$};
\node[circle, blk] at (6.5, 1.4) {\LARGE $x_4$};
\node[circle, blk] at (7.5, 1.5) {\LARGE $\overline{x_4}$};
\end{tikzpicture}
\caption{Arrangement of literal circles for $C_1 = \{x_1, x_2, x_3\}$ in Oredango}
\label{fig:clause1}
\end{figure}

Given an arbitrary 1-in-3SAT instance, we first prepare a row for each clause
and also a column for each of the variables and their negations within that row. 
For each literal of a clause, we place a circle, called a {\it literal circle}, at the cell located at the intersection of the corresponding row and column. 
Hence, for a clause, three literal circles are put in the corresponding row.
We note that the column for a literal is adjacent to that for its negation.
Figure~\ref{fig:clause1} illustrates the arrangement of literal circles for the clause $C_1$.
If a literal circle is black (or white), 
we regard the corresponding literal in the 1-in-3SAT as true (or false), and vice versa. 
Combining the rows corresponding to the clauses and placing the literal circles in the above manner, a specific Oredango instance is reduced from the 1-in-3SAT instance, which is illustrated in Figure~\ref{fig:idea_board}.
Note that the literal circle of $x_1$ differs from that of $\overline{x_1}$.
However, this is not the desired reduction. 
Indeed, the reduced Oredango instance has the following problems. 
\begin{enumerate}
\item[P1:] The colors of the circles that represent the same literal do not necessarily match.
\item[P2:] For a literal circle which is colored black (white), 
its negative literal circle is not necessarily colored white (black).
\item[P3:] 
There exists an Oredango solution that can be interpreted 
as a truth assignment where two of the literals in some clauses are true, 
which is rejected by 1-in-3SAT.
\end{enumerate}

\begin{figure}[hb]
\centering
\begin{tikzpicture}[scale=1.0]
%grid
\draw[step=1.0cm, color={rgb:black, 1;white, 4}] (0.0, 0) grid (8.0, 3.0);
%circles
\node[draw, circle, minimum size=0.7cm, fill=wht] at (0.5, 2.5) {};
\node[draw, circle, minimum size=0.7cm, fill=wht] at (2.5, 2.5) {};
\node[draw, circle, minimum size=0.7cm, fill=wht] at (4.5, 2.5) {};
\node[draw, circle, minimum size=0.7cm, fill=wht] at (1.5, 1.5) {};
\node[draw, circle, minimum size=0.7cm, fill=wht] at (4.5, 1.5) {};
\node[draw, circle, minimum size=0.7cm, fill=wht] at (6.5, 1.5) {};
\node[draw, circle, minimum size=0.7cm, fill=wht] at (2.5, 0.5) {};
\node[draw, circle, minimum size=0.7cm, fill=wht] at (5.5, 0.5) {};
\node[draw, circle, minimum size=0.7cm, fill=wht] at (7.5, 0.5) {};
%variables, clauses
\node[circle, blk] at (-0.5, 2.4) {\LARGE $C_1$};
\node[circle, blk] at (-0.5, 1.4) {\LARGE $C_2$};
\node[circle, blk] at (-0.5, 0.4) {\LARGE $C_3$};
\node[circle, blk] at (0.5, 3.4) {\LARGE $x_1$};
\node[circle, blk] at (1.5, 3.5) {\LARGE $\overline{x_1}$};
\node[circle, blk] at (2.5, 3.4) {\LARGE $x_2$};
\node[circle, blk] at (3.5, 3.5) {\LARGE $\overline{x_2}$};
\node[circle, blk] at (4.5, 3.4) {\LARGE $x_3$};
\node[circle, blk] at (5.5, 3.5) {\LARGE $\overline{x_3}$};
\node[circle, blk] at (6.5, 3.4) {\LARGE $x_4$};
\node[circle, blk] at (7.5, 3.5) {\LARGE $\overline{x_4}$};
\end{tikzpicture}
\caption{Arrangement of the rows corresponding to $C_1, C_2,$ and $C_3$}
\label{fig:idea_board}
\end{figure}
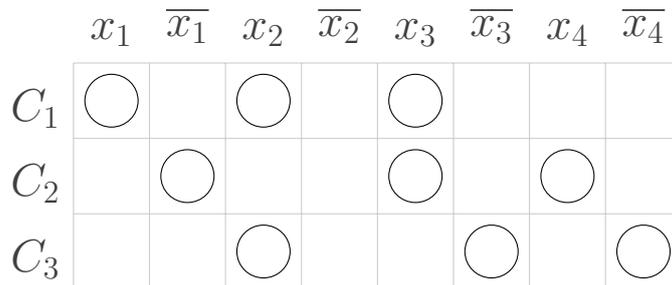

% ---------------- Fig:ガジェットのアイデア ----------------
\begin{figure}[ht]
\centering
\begin{tikzpicture}[scale=0.8]

%grid
\draw[step=0.8cm, color={rgb:black, 1;white, 4}] (0, 0.8) grid (12.8, 10.4);

\node (end) at (0, 1.6) [left] {\large $C_{3}$};
\node (end) at (0, 5.6) [left] {\large $C_{2}$};
\node (end) at (0, 9.6) [left] {\large $C_{1}$};

\node (end) at (1.2, 10.4) [above] {\large $x_{1}$};
\node (end) at (2.0, 10.4) [above] {\large $\overline{x_{1}}$};
\node (end) at (4.4, 10.4) [above] {\large $x_{2}$};
\node (end) at (5.2, 10.4) [above] {\large $\overline{x_{2}}$};
\node (end) at (7.6, 10.4) [above] {\large $x_{3}$};
\node (end) at (8.4, 10.4) [above] {\large $\overline{x_{3}}$};
\node (end) at (10.8, 10.4) [above] {\large $x_{4}$};
\node (end) at (11.6, 10.4) [above] {\large $\overline{x_{4}}$};

%-----------------ガジェットG1-----------------%
\path[pattern=north east lines, pattern color=red] (0.8, 2.4) rectangle + (1.6, 2.4);
\path[pattern=north east lines, pattern color=red] (0.8, 6.4) rectangle + (1.6, 2.4);

\path[pattern=north east lines, pattern color=red] (4.0, 2.4) rectangle + (1.6, 2.4);
\path[pattern=north east lines, pattern color=red] (4.0, 6.4) rectangle + (1.6, 2.4);

\path[pattern=north east lines, pattern color=red] (7.2, 2.4) rectangle + (1.6, 2.4);
\path[pattern=north east lines, pattern color=red] (7.2, 6.4) rectangle + (1.6, 2.4);

\path[pattern=north east lines, pattern color=red] (10.4, 2.4) rectangle + (1.6, 2.4);
\path[pattern=north east lines, pattern color=red] (10.4, 6.4) rectangle + (1.6, 2.4);
%-----------------ガジェットG2-----------------%
\path[pattern=dots, pattern color=blue] (0.0, 1.6) rectangle + (0.8, 0.8);
\path[pattern=dots, pattern color=blue] (0.0, 5.6) rectangle + (0.8, 0.8);
\path[pattern=dots, pattern color=blue] (0.0, 9.6) rectangle + (0.8, 0.8);

\path[pattern=dots, pattern color=blue] (2.4, 1.6) rectangle + (1.6, 0.8);
\path[pattern=dots, pattern color=blue] (2.4, 5.6) rectangle + (1.6, 0.8);
\path[pattern=dots, pattern color=blue] (2.4, 9.6) rectangle + (1.6, 0.8);

\path[pattern=dots, pattern color=blue] (5.6, 1.6) rectangle + (1.6, 0.8);
\path[pattern=dots, pattern color=blue] (5.6, 5.6) rectangle + (1.6, 0.8);
\path[pattern=dots, pattern color=blue] (5.6, 9.6) rectangle + (1.6, 0.8);

\path[pattern=dots, pattern color=blue] (8.8, 1.6) rectangle + (1.6, 0.8);
\path[pattern=dots, pattern color=blue] (8.8, 5.6) rectangle + (1.6, 0.8);
\path[pattern=dots, pattern color=blue] (8.8, 9.6) rectangle + (1.6, 0.8);

\path[pattern=dots, pattern color=blue] (12.0, 1.6) rectangle + (0.8, 0.8);
\path[pattern=dots, pattern color=blue] (12.0, 5.6) rectangle + (0.8, 0.8);
\path[pattern=dots, pattern color=blue] (12.0, 9.6) rectangle + (0.8, 0.8);

\path[pattern=dots, pattern color=blue] (0.0, 0.8) rectangle + (12.8, 0.8);
\path[pattern=dots, pattern color=blue] (0.0, 4.8) rectangle + (12.8, 0.8);
\path[pattern=dots, pattern color=blue] (0.0, 8.8) rectangle + (12.8, 0.8);
%-----------------ガジェットG3-----------------%
\path[pattern=vertical lines, pattern color=green] (0.0, 2.4) rectangle + (0.8, 2.4);
\path[pattern=vertical lines, pattern color=green] (0.0, 6.4) rectangle + (0.8, 2.4);

\path[pattern=vertical lines, pattern color=green] (2.4, 2.4) rectangle + (1.6, 2.4);
\path[pattern=vertical lines, pattern color=green] (2.4, 6.4) rectangle + (1.6, 2.4);

\path[pattern=vertical lines, pattern color=green] (5.6, 2.4) rectangle + (1.6, 2.4);
\path[pattern=vertical lines, pattern color=green] (5.6, 6.4) rectangle + (1.6, 2.4);

\path[pattern=vertical lines, pattern color=green] (8.8, 2.4) rectangle + (1.6, 2.4);
\path[pattern=vertical lines, pattern color=green] (8.8, 6.4) rectangle + (1.6, 2.4);

\path[pattern=vertical lines, pattern color=green] (12.0, 2.4) rectangle + (0.8, 2.4);
\path[pattern=vertical lines, pattern color=green] (12.0, 6.4) rectangle + (0.8, 2.4);
%-----------------ガジェットG1-----------------%
\node (end) at (1.1, 3.6) [right, font=\large] {G1};
\node (end) at (1.1, 7.6) [right, font=\large] {G1};

\node (end) at (4.3, 3.6) [right, font=\large] {G1};
\node (end) at (4.3, 7.6) [right, font=\large] {G1};

\node (end) at (7.5, 3.6) [right, font=\large] {G1};
\node (end) at (7.5, 7.6) [right, font=\large] {G1};

\node (end) at (10.7, 3.6) [right, font=\large] {G1};
\node (end) at (10.7, 7.6) [right, font=\large] {G1};
%-----------------ガジェットG2-----------------%
\node (end) at (5.9, 1.2) [right, font=\large] {G2};
\node (end) at (5.9, 5.2) [right, font=\large] {G2};
\node (end) at (5.9, 9.2) [right, font=\large] {G2};
%-----------------ガジェットG3-----------------%
\node (end) at (2.7, 3.6) [right, font=\large] {G3};
\node (end) at (2.7, 7.6) [right, font=\large] {G3};

\node (end) at (5.9, 3.6) [right, font=\large] {G3};
\node (end) at (5.9, 7.6) [right, font=\large] {G3};

\node (end) at (9.1, 3.6) [right, font=\large] {G3};
\node (end) at (9.1, 7.6) [right, font=\large] {G3};

%C_1
\node[draw, circle, minimum size=0.5cm, fill=wht] at (1.2, 10.0) {};
\node[draw, circle, minimum size=0.5cm, fill=wht] at (4.4, 10.0) {};
\node[draw, circle, minimum size=0.5cm, fill=wht] at (7.6, 10.0) {};
%C2
\node[draw, circle, minimum size=0.5cm, fill=wht] at (2.0, 6.0) {};
\node[draw, circle, minimum size=0.5cm, fill=wht] at (7.6, 6.0) {};
\node[draw, circle, minimum size=0.5cm, fill=wht] at (10.8, 6.0) {};
%C3
\node[draw, circle, minimum size=0.5cm, fill=wht] at (4.4, 2.0) {};
\node[draw, circle, minimum size=0.5cm, fill=wht] at (8.4, 2.0) {};
\node[draw, circle, minimum size=0.5cm, fill=wht] at (11.6, 2.0) {};

\draw[line width=0.6mm] (0, 0.8) -- (12.8, 0.8);
\draw[line width=0.6mm] (0, 2.4) -- (12.8, 2.4);
\draw[line width=0.6mm] (0, 4.8) -- (12.8, 4.8);
\draw[line width=0.6mm] (0, 6.4) -- (12.8, 6.4);

\draw[line width=0.6mm] (0, 8.8) -- (12.8, 8.8);
\draw[line width=0.6mm] (0, 10.4) -- (12.8, 10.4);

\draw[line width=0.6mm] (0, 0.8) -- (0, 2.4);
\draw[line width=0.6mm] (0, 4.8) -- (0, 6.4);
\draw[line width=0.6mm] (0, 8.8) -- (0, 10.4);

\draw[line width=0.6mm] (12.8, 0.8) -- (12.8, 2.4);
\draw[line width=0.6mm] (12.8, 4.8) -- (12.8, 6.4);
\draw[line width=0.6mm] (12.8, 8.8) -- (12.8, 10.4);
\end{tikzpicture}
\caption{Idea of constructing the board of Oredango from 1-in-3SAT}
\label{fig:idea}
\end{figure}
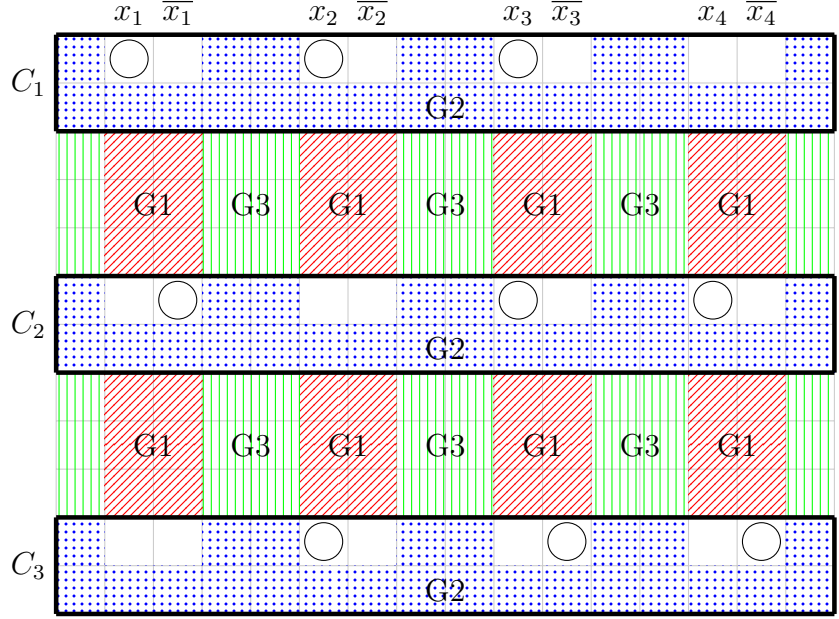

To resolve these problems, we extend the board as illustrated in Figure~\ref{fig:idea}, where the shaded areas G1, G2, and G3, called {\it gadgets}, are newly added.  
In these gadgets, circles and skewers will be placed in a certain manner explained later.
The roles of the gadgets G1, G2, and G3 are as follows:  
\begin{enumerate}
\item[G1:] 
Gadget to adjust the colors of the literal circles for the same literal.
This is for P1 and P2.
\item[G2:] Gadget to adjust so that exactly one of the literal circles in each clause is true.
This is for P3.
\item[G3:] With only G1 and G2, the obtained Oredango is necessarily infeasible.
G3 serves as a gadget to resolve this infeasibility.
\end{enumerate}
In what follows, we precisely explain the gadgets and the manner of adding circles and skews.

\subsection{Arrangement of components in gadget G1} \label{sec:G1}
\begin{figure}[hb]
\centering
\begin{tikzpicture}[scale=1.0]
\draw[step=1.0cm, color={rgb:black, 1;white, 4}] (0.0, 0.0) grid (3.0, 2.0);
\draw[line width=0.6mm] (0.5, 0.5) -- (1.5, 1.5);
\draw[line width=0.6mm] (0.5, 1.5) -- (1.5, 0.5);
\node[draw, circle, minimum size=0.7cm, fill=wht] at (0.5, 1.5) {};
\node[draw, circle, minimum size=0.7cm, fill=wht] at (1.5, 1.5) {};
\node[draw, circle, minimum size=0.7cm, fill=wht] at (0.5, 0.5) {};
\node[draw, circle, minimum size=0.7cm, fill=wht] at (1.5, 0.5) {};
\node[draw, circle, minimum size=0.7cm, fill=wht] at (2.5, 1.5) {};
\node[draw, circle, minimum size=0.7cm, fill=wht] at (2.5, 0.5) {};
\node[circle, blk] at (0.5, 0.5) {\LARGE $1$};
\node[circle, blk] at (0.5, 1.5) {\LARGE $1$};
\node[circle, blk] at (2.5, 1.5) {\LARGE $1$};
\node[circle, blk] at (2.5, 0.5) {\LARGE $1$};
\node[circle, black] at (1.5, -0.5) {\LARGE (i)};
\end{tikzpicture}
\hspace{0.7cm}
\begin{tikzpicture}[scale=1.0]
\draw[step=1.0cm, color={rgb:black, 1;white, 4}] (0.0, 0.0) grid (3.0, 2.0);
\draw[line width=0.6mm] (0.5, 0.5) -- (1.5, 1.5);
\draw[line width=0.6mm] (0.5, 1.5) -- (1.5, 0.5);
\node[draw, circle, minimum size=0.7cm, fill=blk] at (0.5, 1.5) {};
\node[draw, circle, minimum size=0.7cm, fill=wht] at (1.5, 1.5) {};
\node[draw, circle, minimum size=0.7cm, fill=blk] at (0.5, 0.5) {};
\node[draw, circle, minimum size=0.7cm, fill=wht] at (1.5, 0.5) {};
\node[draw, circle, minimum size=0.7cm, fill=blk] at (2.5, 1.5) {};
\node[draw, circle, minimum size=0.7cm, fill=blk] at (2.5, 0.5) {};
\node[circle, wht] at (0.5, 0.5) {\LARGE $1$};
\node[circle, wht] at (0.5, 1.5) {\LARGE $1$};
\node[circle, wht] at (2.5, 1.5) {\LARGE $1$};
\node[circle, wht] at (2.5, 0.5) {\LARGE $1$};
\node[circle, black] at (3.5, 1.0) {\LARGE or};
\draw[step=1.0cm, color={rgb:black, 1;white, 4}] (4.0, 0.0) grid (7.0, 2.0);
\draw[line width=0.6mm] (4.5, 0.5) -- (5.5, 1.5);
\draw[line width=0.6mm] (4.5, 1.5) -- (5.5, 0.5);
\node[draw, circle, minimum size=0.7cm, fill=wht] at (4.5, 1.5) {};
\node[draw, circle, minimum size=0.7cm, fill=blk] at (5.5, 1.5) {};
\node[draw, circle, minimum size=0.7cm, fill=wht] at (4.5, 0.5) {};
\node[draw, circle, minimum size=0.7cm, fill=blk] at (5.5, 0.5) {};
\node[draw, circle, minimum size=0.7cm, fill=blk] at (6.5, 1.5) {};
\node[draw, circle, minimum size=0.7cm, fill=blk] at (6.5, 0.5) {};
\node[circle, blk] at (4.5, 0.5) {\LARGE $1$};
\node[circle, blk] at (4.5, 1.5) {\LARGE $1$};
\node[circle, wht] at (6.5, 1.5) {\LARGE $1$};
\node[circle, wht] at (6.5, 0.5) {\LARGE $1$};
\node[circle, black] at (3.5, -0.5) {\LARGE (ii)};
\end{tikzpicture}
\caption{Circles and skewers placed in G1 (i), and all colored patterns (ii)}
\label{fig:G1_circles_skewers}
\end{figure}
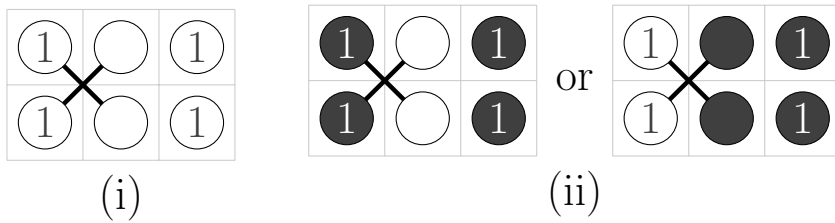

This gadget is for resolving P1.
According to the rules, the circles corresponding to the identical literal must be colored identically, and the color of a literal circle must be different from that of its negation.
Nonetheless, this may not be the case at the stage of the idea in Figure~\ref{fig:idea_board}.
To resolve these inconsistencies, 
we place circles, skewers, and integers for the red shaded area in Figure~\ref{fig:idea}.
In that configuration, we make use of the key pattern shown in Figure~\ref{fig:G1_circles_skewers}.
Note that, for the input in Figure~\ref{fig:G1_circles_skewers}(i),
the possible coloring manners are only two shown in Figure~\ref{fig:G1_circles_skewers}(ii).

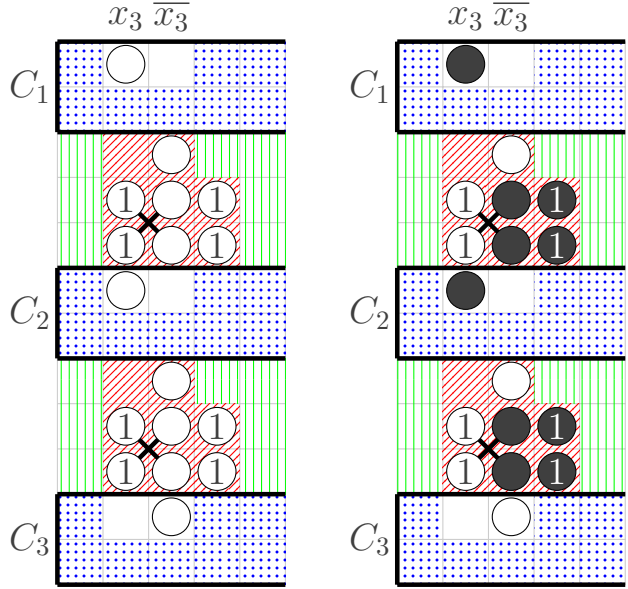
\begin{figure}[hb]
\centering
%-----------------------(left)-----------------------
\begin{tikzpicture}[scale=0.6]
%grid
\draw[step=1.0cm, color={rgb:black, 1;white, 4}] (-1.0, -6.0) grid (4.0, 6.0);
%Gadget
\path[pattern=north east lines, pattern color=red] (0.0, 1.0) rectangle + (2.0, 3.0);
\path[pattern=north east lines, pattern color=red] (0.0, -4.0) rectangle + (2.0, 3.0);
\path[pattern=north east lines, pattern color=red] (2.0, 1.0) rectangle + (1.0, 2.0);
\path[pattern=north east lines, pattern color=red] (2.0, -4.0) rectangle + (1.0, 2.0);
\path[pattern=dots, pattern color=blue] (-1.0, 5.0) rectangle + (1.0, 1.0);
\path[pattern=dots, pattern color=blue] (2.0, 5.0) rectangle + (2.0, 1.0);
\path[pattern=dots, pattern color=blue] (-1.0, 4.0) rectangle + (5.0, 1.0);
\path[pattern=dots, pattern color=blue] (-1.0, 0.0) rectangle + (1.0, 1.0);
\path[pattern=dots, pattern color=blue] (2.0, 0.0) rectangle + (2.0, 1.0);
\path[pattern=dots, pattern color=blue] (-1.0, -1.0) rectangle + (5.0, 1.0);
\path[pattern=dots, pattern color=blue] (-1.0, -5.0) rectangle + (1.0, 1.0);
\path[pattern=dots, pattern color=blue] (2.0, -5.0) rectangle + (2.0, 1.0);
\path[pattern=dots, pattern color=blue] (-1.0, -6.0) rectangle + (5.0, 1.0);
\path[pattern=vertical lines, pattern color=green] (-1.0, 1.0) rectangle + (1.0, 3.0);
\path[pattern=vertical lines, pattern color=green] (-1.0, -4.0) rectangle + (1.0, 3.0);
\path[pattern=vertical lines, pattern color=green] (3.0, 1.0) rectangle + (1.0, 3.0);
\path[pattern=vertical lines, pattern color=green] (2.0, 3.0) rectangle + (1.0, 1.0);
\path[pattern=vertical lines, pattern color=green] (3.0, -4.0) rectangle + (1.0, 3.0);
\path[pattern=vertical lines, pattern color=green] (2.0, -2.0) rectangle + (1.0, 1.0);
%skewers
\draw[line width=0.6mm] (0.5, 1.5) -- (1.5, 2.5);
\draw[line width=0.6mm] (0.5, 2.5) -- (1.5, 1.5);
\draw[line width=0.6mm] (0.5, -2.5) -- (1.5, -3.5);
\draw[line width=0.6mm] (0.5, -3.5) -- (1.5, -2.5);
%circles
\node[draw, circle, minimum size=0.5cm, fill=wht] at (0.5, 5.5) {};
\node[draw, circle, minimum size=0.5cm, fill=wht] at (1.5, 3.5) {};
\node[draw, circle, minimum size=0.5cm, fill=wht] at (0.5, 2.5) {};
\node[draw, circle, minimum size=0.5cm, fill=wht] at (1.5, 2.5) {};
\node[draw, circle, minimum size=0.5cm, fill=wht] at (0.5, 1.5) {};
\node[draw, circle, minimum size=0.5cm, fill=wht] at (1.5, 1.5) {};
\node[draw, circle, minimum size=0.5cm, fill=wht] at (0.5, 0.5) {};
\node[draw, circle, minimum size=0.5cm, fill=wht] at (1.5, -1.5) {};
\node[draw, circle, minimum size=0.5cm, fill=wht] at (0.5, -2.5) {};
\node[draw, circle, minimum size=0.5cm, fill=wht] at (1.5, -2.5) {};
\node[draw, circle, minimum size=0.5cm, fill=wht] at (0.5, -3.5) {};
\node[draw, circle, minimum size=0.5cm, fill=wht] at (1.5, -3.5) {};
\node[draw, circle, minimum size=0.5cm, fill=wht] at (1.5, -4.5) {};
\node[draw, circle, minimum size=0.5cm, fill=wht] at (2.5, 2.5) {};
\node[draw, circle, minimum size=0.5cm, fill=wht] at (2.5, 1.5) {};
\node[draw, circle, minimum size=0.5cm, fill=wht] at (2.5, -2.5) {};
\node[draw, circle, minimum size=0.5cm, fill=wht] at (2.5, -3.5) {};
%variables, clauses
\node[circle, blk] at (0.5, 6.5) {\Large $x_3$};
\node[circle, blk] at (1.5, 6.57) {\Large $\overline{x_3}$};
\node[circle, blk] at (-1.6, 5.0) {\Large $C_1$};
\node[circle, blk] at (-1.6, 0.0) {\Large $C_2$};
\node[circle, blk] at (-1.6, -5.0) {\Large $C_3$};
\node[circle, blk] at (0.5, 1.5) {\Large $1$};
\node[circle, blk] at (0.5, 2.5) {\Large $1$};
\node[circle, blk] at (0.5, -2.5) {\Large $1$};
\node[circle, blk] at (0.5, -3.5) {\Large $1$};
\node[circle, blk] at (2.5, 2.5) {\Large $1$};
\node[circle, blk] at (2.5, 1.5) {\Large $1$};
\node[circle, blk] at (2.5, -2.5) {\Large $1$};
\node[circle, blk] at (2.5, -3.5) {\Large $1$};
%frame
\draw[line width=0.6mm] (-1.0, 4.0) -- (4.0, 4.0);
\draw[line width=0.6mm] (-1.0, 4.0) -- (-1.0, 6.0);
\draw[line width=0.6mm] (-1.0, 6.0) -- (4.0, 6.0);
\draw[line width=0.6mm] (-1.0, 1.0) -- (4.0, 1.0);
\draw[line width=0.6mm] (-1.0, 1.0) -- (-1.0, -1.0);
\draw[line width=0.6mm] (-1.0, -1.0) -- (4.0, -1.0);
\draw[line width=0.6mm] (-1.0, -4.0) -- (4.0, -4.0);
\draw[line width=0.6mm] (-1.0, -4.0) -- (-1.0, -6.0);
\draw[line width=0.6mm] (-1.0, -6.0) -- (4.0, -6.0);
\end{tikzpicture}
\hspace{0.3cm}
%-----------------------(right)-----------------------
\begin{tikzpicture}[scale=0.6]
%grid
\draw[step=1.0cm, color={rgb:black, 1;white, 4}] (-1.0, -6.0) grid (4.0, 6.0);
%Gadget
\path[pattern=north east lines, pattern color=red] (0.0, 1.0) rectangle + (2.0, 3.0);
\path[pattern=north east lines, pattern color=red] (0.0, -4.0) rectangle + (2.0, 3.0);
\path[pattern=north east lines, pattern color=red] (2.0, 1.0) rectangle + (1.0, 2.0);
\path[pattern=north east lines, pattern color=red] (2.0, -4.0) rectangle + (1.0, 2.0);
\path[pattern=dots, pattern color=blue] (-1.0, 5.0) rectangle + (1.0, 1.0);
\path[pattern=dots, pattern color=blue] (2.0, 5.0) rectangle + (2.0, 1.0);
\path[pattern=dots, pattern color=blue] (-1.0, 4.0) rectangle + (5.0, 1.0);
\path[pattern=dots, pattern color=blue] (-1.0, 0.0) rectangle + (1.0, 1.0);
\path[pattern=dots, pattern color=blue] (2.0, 0.0) rectangle + (2.0, 1.0);
\path[pattern=dots, pattern color=blue] (-1.0, -1.0) rectangle + (5.0, 1.0);
\path[pattern=dots, pattern color=blue] (-1.0, -5.0) rectangle + (1.0, 1.0);
\path[pattern=dots, pattern color=blue] (2.0, -5.0) rectangle + (2.0, 1.0);
\path[pattern=dots, pattern color=blue] (-1.0, -6.0) rectangle + (5.0, 1.0);
\path[pattern=vertical lines, pattern color=green] (-1.0, 1.0) rectangle + (1.0, 3.0);
\path[pattern=vertical lines, pattern color=green] (-1.0, -4.0) rectangle + (1.0, 3.0);
\path[pattern=vertical lines, pattern color=green] (3.0, 1.0) rectangle + (1.0, 3.0);
\path[pattern=vertical lines, pattern color=green] (2.0, 3.0) rectangle + (1.0, 1.0);
\path[pattern=vertical lines, pattern color=green] (3.0, -4.0) rectangle + (1.0, 3.0);
\path[pattern=vertical lines, pattern color=green] (2.0, -2.0) rectangle + (1.0, 1.0);
%skewers
\draw[line width=0.6mm] (0.5, 1.5) -- (1.5, 2.5);
\draw[line width=0.6mm] (0.5, 2.5) -- (1.5, 1.5);
\draw[line width=0.6mm] (0.5, -2.5) -- (1.5, -3.5);
\draw[line width=0.6mm] (0.5, -3.5) -- (1.5, -2.5);
%circles
\node[draw, circle, minimum size=0.5cm, fill=blk] at (0.5, 5.5) {};
\node[draw, circle, minimum size=0.5cm, fill=wht] at (1.5, 3.5) {};
\node[draw, circle, minimum size=0.5cm, fill=wht] at (0.5, 2.5) {};
\node[draw, circle, minimum size=0.5cm, fill=blk] at (1.5, 2.5) {};
\node[draw, circle, minimum size=0.5cm, fill=wht] at (0.5, 1.5) {};
\node[draw, circle, minimum size=0.5cm, fill=blk] at (1.5, 1.5) {};
\node[draw, circle, minimum size=0.5cm, fill=blk] at (0.5, 0.5) {};
\node[draw, circle, minimum size=0.5cm, fill=wht] at (1.5, -1.5) {};
\node[draw, circle, minimum size=0.5cm, fill=wht] at (0.5, -2.5) {};
\node[draw, circle, minimum size=0.5cm, fill=blk] at (1.5, -2.5) {};
\node[draw, circle, minimum size=0.5cm, fill=wht] at (0.5, -3.5) {};
\node[draw, circle, minimum size=0.5cm, fill=blk] at (1.5, -3.5) {};
\node[draw, circle, minimum size=0.5cm, fill=wht] at (1.5, -4.5) {};
\node[draw, circle, minimum size=0.5cm, fill=blk] at (2.5, 2.5) {};
\node[draw, circle, minimum size=0.5cm, fill=blk] at (2.5, 1.5) {};
\node[draw, circle, minimum size=0.5cm, fill=blk] at (2.5, -2.5) {};
\node[draw, circle, minimum size=0.5cm, fill=blk] at (2.5, -3.5) {};
%variables, clauses
\node[circle, blk] at (0.5, 6.5) {\Large $x_3$};
\node[circle, blk] at (1.5, 6.57) {\Large $\overline{x_3}$};
\node[circle, blk] at (-1.6, 5.0) {\Large $C_1$};
\node[circle, blk] at (-1.6, 0.0) {\Large $C_2$};
\node[circle, blk] at (-1.6, -5.0) {\Large $C_3$};
\node[circle, blk] at (0.5, 1.5) {\Large $1$};
\node[circle, blk] at (0.5, 2.5) {\Large $1$};
\node[circle, blk] at (0.5, -2.5) {\Large $1$};
\node[circle, blk] at (0.5, -3.5) {\Large $1$};
\node[circle, wht] at (2.5, 2.5) {\Large $1$};
\node[circle, wht] at (2.5, 1.5) {\Large $1$};
\node[circle, wht] at (2.5, -2.5) {\Large $1$};
\node[circle, wht] at (2.5, -3.5) {\Large $1$};
%frame
\draw[line width=0.6mm] (-1.0, 4.0) -- (4.0, 4.0);
\draw[line width=0.6mm] (-1.0, 4.0) -- (-1.0, 6.0);
\draw[line width=0.6mm] (-1.0, 6.0) -- (4.0, 6.0);
\draw[line width=0.6mm] (-1.0, 1.0) -- (4.0, 1.0);
\draw[line width=0.6mm] (-1.0, 1.0) -- (-1.0, -1.0);
\draw[line width=0.6mm] (-1.0, -1.0) -- (4.0, -1.0);
\draw[line width=0.6mm] (-1.0, -4.0) -- (4.0, -4.0);
\draw[line width=0.6mm] (-1.0, -4.0) -- (-1.0, -6.0);
\draw[line width=0.6mm] (-1.0, -6.0) -- (4.0, -6.0);
\end{tikzpicture}
\caption{Literal circles for $x_3$ and $\overline{x_3}$ with G1 (left), and colored board corresponding to $x_3 = 1 (\overline{x_3}=0)$ (right)}
\label{G1_model}
\end{figure}

We adjust the colors of literal circles using the key pattern together with adding circles, called {\it support circles}.
Figure~\ref{G1_model}(left) illustrates the proposed configuration on gadget G1, 
and Figure~\ref{G1_model}(right) shows the color pattern in the case of setting $x_3 = 1$.
In the same figure,
at the column corresponding to the negation of the literal we add support circles 
below the rows in which the literal circles are placed.
For example, for the literal circle representing $x_3$ placed in clause $C_1$, 
its support circle is placed in the third row and the third column from the top. 

This configuration can resolve problems P1 and P2 from the following fact:
From rule (d), when two black (white) circles are placed consecutively in the same column, 
the color of the circles sandwiching these two black (white) circles must be white (black). 
Therefore, in a column where two literal circles are present, placing two additional circles of the same color between the two literal circles forces them to be colored identically, but differently from the placed circles.
Indeed, in Figure~\ref{G1_model}(left), 
all the literal circles representing $x_3$ are colored the same,
furthermore, all the literal circles representing $\overline{x_3}$ are always colored differently from $x_3$.

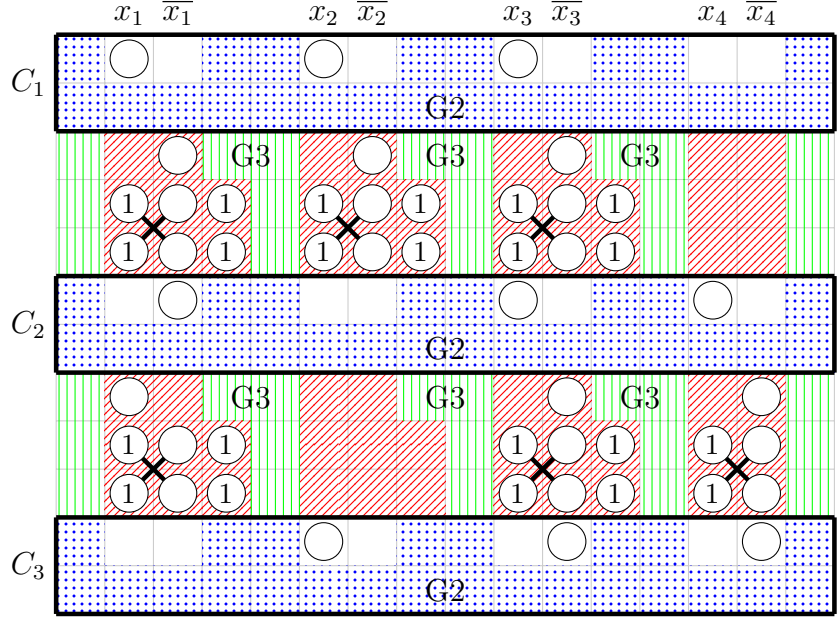
\begin{figure}[ht]
\centering
\begin{tikzpicture}[scale=0.8]

%grid
\draw[step=0.8cm, color={rgb:black, 1;white, 4}] (0, 0.8) grid (12.8, 10.4);

\node (end) at (0, 1.6) [left] {\large $C_{3}$};
\node (end) at (0, 5.6) [left] {\large $C_{2}$};
\node (end) at (0, 9.6) [left] {\large $C_{1}$};

\node (end) at (1.2, 10.4) [above] {\large $x_{1}$};
\node (end) at (2.0, 10.4) [above] {\large $\overline{x_{1}}$};
\node (end) at (4.4, 10.4) [above] {\large $x_{2}$};
\node (end) at (5.2, 10.4) [above] {\large $\overline{x_{2}}$};
\node (end) at (7.6, 10.4) [above] {\large $x_{3}$};
\node (end) at (8.4, 10.4) [above] {\large $\overline{x_{3}}$};
\node (end) at (10.8, 10.4) [above] {\large $x_{4}$};
\node (end) at (11.6, 10.4) [above] {\large $\overline{x_{4}}$};

%-----------------ガジェットG1-----------------%
\path[pattern=north east lines, pattern color=red] (0.8, 2.4) rectangle + (1.6, 2.4);
\path[pattern=north east lines, pattern color=red] (2.4, 2.4) rectangle + (0.8, 1.6);
\path[pattern=north east lines, pattern color=red] (0.8, 6.4) rectangle + (1.6, 2.4);
\path[pattern=north east lines, pattern color=red] (2.4, 6.4) rectangle + (0.8, 1.6);

\path[pattern=north east lines, pattern color=red] (4.0, 2.4) rectangle + (1.6, 2.4);
\path[pattern=north east lines, pattern color=red] (5.6, 2.4) rectangle + (0.8, 1.6);
\path[pattern=north east lines, pattern color=red] (4.0, 6.4) rectangle + (1.6, 2.4);
\path[pattern=north east lines, pattern color=red] (5.6, 6.4) rectangle + (0.8, 1.6);

\path[pattern=north east lines, pattern color=red] (7.2, 2.4) rectangle + (1.6, 2.4);
\path[pattern=north east lines, pattern color=red] (8.8, 2.4) rectangle + (0.8, 1.6);
\path[pattern=north east lines, pattern color=red] (7.2, 6.4) rectangle + (1.6, 2.4);
\path[pattern=north east lines, pattern color=red] (8.8, 6.4) rectangle + (0.8, 1.6);

\path[pattern=north east lines, pattern color=red] (10.4, 2.4) rectangle + (1.6, 2.4);
\path[pattern=north east lines, pattern color=red] (10.4, 6.4) rectangle + (1.6, 2.4);
%-----------------ガジェットG2-----------------%
\path[pattern=dots, pattern color=blue] (0.0, 1.6) rectangle + (0.8, 0.8);
\path[pattern=dots, pattern color=blue] (0.0, 5.6) rectangle + (0.8, 0.8);
\path[pattern=dots, pattern color=blue] (0.0, 9.6) rectangle + (0.8, 0.8);

\path[pattern=dots, pattern color=blue] (2.4, 1.6) rectangle + (1.6, 0.8);
\path[pattern=dots, pattern color=blue] (2.4, 5.6) rectangle + (1.6, 0.8);
\path[pattern=dots, pattern color=blue] (2.4, 9.6) rectangle + (1.6, 0.8);

\path[pattern=dots, pattern color=blue] (5.6, 1.6) rectangle + (1.6, 0.8);
\path[pattern=dots, pattern color=blue] (5.6, 5.6) rectangle + (1.6, 0.8);
\path[pattern=dots, pattern color=blue] (5.6, 9.6) rectangle + (1.6, 0.8);

\path[pattern=dots, pattern color=blue] (8.8, 1.6) rectangle + (1.6, 0.8);
\path[pattern=dots, pattern color=blue] (8.8, 5.6) rectangle + (1.6, 0.8);
\path[pattern=dots, pattern color=blue] (8.8, 9.6) rectangle + (1.6, 0.8);

\path[pattern=dots, pattern color=blue] (12.0, 1.6) rectangle + (0.8, 0.8);
\path[pattern=dots, pattern color=blue] (12.0, 5.6) rectangle + (0.8, 0.8);
\path[pattern=dots, pattern color=blue] (12.0, 9.6) rectangle + (0.8, 0.8);

\path[pattern=dots, pattern color=blue] (0.0, 0.8) rectangle + (12.8, 0.8);
\path[pattern=dots, pattern color=blue] (0.0, 4.8) rectangle + (12.8, 0.8);
\path[pattern=dots, pattern color=blue] (0.0, 8.8) rectangle + (12.8, 0.8);
%-----------------ガジェットG3-----------------%
\path[pattern=vertical lines, pattern color=green] (0.0, 2.4) rectangle + (0.8, 2.4);
\path[pattern=vertical lines, pattern color=green] (0.0, 6.4) rectangle + (0.8, 2.4);

\path[pattern=vertical lines, pattern color=green] (3.2, 2.4) rectangle + (0.8, 1.6);
\path[pattern=vertical lines, pattern color=green] (2.4, 4.0) rectangle + (1.6, 0.8);
\path[pattern=vertical lines, pattern color=green] (3.2, 6.4) rectangle + (0.8, 1.6);
\path[pattern=vertical lines, pattern color=green] (2.4, 8.0) rectangle + (1.6, 0.8);

\path[pattern=vertical lines, pattern color=green] (6.4, 2.4) rectangle + (0.8, 1.6);
\path[pattern=vertical lines, pattern color=green] (5.6, 4.0) rectangle + (1.6, 0.8);
\path[pattern=vertical lines, pattern color=green] (6.4, 6.4) rectangle + (0.8, 1.6);
\path[pattern=vertical lines, pattern color=green] (5.6, 8.0) rectangle + (1.6, 0.8);

\path[pattern=vertical lines, pattern color=green] (9.6, 2.4) rectangle + (0.8, 1.6);
\path[pattern=vertical lines, pattern color=green] (8.8, 4.0) rectangle + (1.6, 0.8);
\path[pattern=vertical lines, pattern color=green] (9.6, 6.4) rectangle + (0.8, 1.6);
\path[pattern=vertical lines, pattern color=green] (8.8, 8.0) rectangle + (1.6, 0.8);

\path[pattern=vertical lines, pattern color=green] (12.0, 2.4) rectangle + (0.8, 2.4);
\path[pattern=vertical lines, pattern color=green] (12.0, 6.4) rectangle + (0.8, 2.4);
%-----------串の描画-----------
\draw[line width=0.6mm] (1.2, 6.8) -- (2.0, 7.6);
\draw[line width=0.6mm] (1.2, 7.6) -- (2.0, 6.8);
\draw[line width=0.6mm] (1.2, 2.8) -- (2.0, 3.6);
\draw[line width=0.6mm] (1.2, 3.6) -- (2.0, 2.8);

\draw[line width=0.6mm] (4.4, 6.8) -- (5.2, 7.6);
\draw[line width=0.6mm] (4.4, 7.6) -- (5.2, 6.8);

\draw[line width=0.6mm] (7.6, 6.8) -- (8.4, 7.6);
\draw[line width=0.6mm] (7.6, 7.6) -- (8.4, 6.8);
\draw[line width=0.6mm] (7.6, 2.8) -- (8.4, 3.6);
\draw[line width=0.6mm] (7.6, 3.6) -- (8.4, 2.8);

\draw[line width=0.6mm] (10.8, 2.8) -- (11.6, 3.6);
\draw[line width=0.6mm] (10.8, 3.6) -- (11.6, 2.8);
%-----------------ガジェットG3-----------------%
\node (end) at (5.9, 1.2) [right, font=\large] {G2};
\node (end) at (5.9, 5.2) [right, font=\large] {G2};
\node (end) at (5.9, 9.2) [right, font=\large] {G2};
%-----------------ガジェットG2-----------------%
\node (end) at (2.7, 4.4) [right, font=\large] {G3};
\node (end) at (2.7, 8.4) [right, font=\large] {G3};

\node (end) at (5.9, 4.4) [right, font=\large] {G3};
\node (end) at (5.9, 8.4) [right, font=\large] {G3};

\node (end) at (9.1, 4.4) [right, font=\large] {G3};
\node (end) at (9.1, 8.4) [right, font=\large] {G3};
%-----------だんごの描画-----------
%C_1
\node[draw, circle, minimum size=0.5cm, fill=wht] at (1.2, 10.0) {};
\node[draw, circle, minimum size=0.5cm, fill=wht] at (2.0, 8.4) {};
\node[draw, circle, minimum size=0.5cm, fill=wht] at (4.4, 10.0) {};
\node[draw, circle, minimum size=0.5cm, fill=wht] at (5.2, 8.4) {};
\node[draw, circle, minimum size=0.5cm, fill=wht] at (7.6, 10.0) {};
\node[draw, circle, minimum size=0.5cm, fill=wht] at (8.4, 8.4) {};
%C2
\node[draw, circle, minimum size=0.5cm, fill=wht] at (1.2, 4.4) {};
\node[draw, circle, minimum size=0.5cm, fill=wht] at (2.0, 6.0) {};
\node[draw, circle, minimum size=0.5cm, fill=wht] at (7.6, 6.0) {};
\node[draw, circle, minimum size=0.5cm, fill=wht] at (8.4, 4.4) {};
\node[draw, circle, minimum size=0.5cm, fill=wht] at (10.8, 6.0) {};
\node[draw, circle, minimum size=0.5cm, fill=wht] at (11.6, 4.4) {};
%C3
\node[draw, circle, minimum size=0.5cm, fill=wht] at (4.4, 2.0) {};
\node[draw, circle, minimum size=0.5cm, fill=wht] at (8.4, 2.0) {};
\node[draw, circle, minimum size=0.5cm, fill=wht] at (11.6, 2.0) {};
%--------G1--------
%----x1----
\node[draw, circle, minimum size=0.5cm, fill=wht] at (1.2, 7.6) {};
\node[draw, circle, minimum size=0.5cm, fill=wht] at (2.0, 7.6) {};
\node[draw, circle, minimum size=0.5cm, fill=wht] at (2.8, 7.6) {};
\node[draw, circle, minimum size=0.5cm, fill=wht] at (1.2, 6.8) {};
\node[draw, circle, minimum size=0.5cm, fill=wht] at (2.0, 6.8) {};
\node[draw, circle, minimum size=0.5cm, fill=wht] at (2.8, 6.8) {};
\node[draw, circle, minimum size=0.5cm, fill=wht] at (1.2, 3.6) {};
\node[draw, circle, minimum size=0.5cm, fill=wht] at (2.0, 3.6) {};
\node[draw, circle, minimum size=0.5cm, fill=wht] at (2.8, 3.6) {};
\node[draw, circle, minimum size=0.5cm, fill=wht] at (1.2, 2.8) {};
\node[draw, circle, minimum size=0.5cm, fill=wht] at (2.0, 2.8) {};
\node[draw, circle, minimum size=0.5cm, fill=wht] at (2.8, 2.8) {};
%----x2----
\node[draw, circle, minimum size=0.5cm, fill=wht] at (4.4, 7.6) {};
\node[draw, circle, minimum size=0.5cm, fill=wht] at (5.2, 7.6) {};
\node[draw, circle, minimum size=0.5cm, fill=wht] at (6.0, 7.6) {};
\node[draw, circle, minimum size=0.5cm, fill=wht] at (4.4, 6.8) {};
\node[draw, circle, minimum size=0.5cm, fill=wht] at (5.2, 6.8) {};
\node[draw, circle, minimum size=0.5cm, fill=wht] at (6.0, 6.8) {};
%----x3----
\node[draw, circle, minimum size=0.5cm, fill=wht] at (7.6, 7.6) {};
\node[draw, circle, minimum size=0.5cm, fill=wht] at (8.4, 7.6) {};
\node[draw, circle, minimum size=0.5cm, fill=wht] at (9.2, 7.6) {};
\node[draw, circle, minimum size=0.5cm, fill=wht] at (7.6, 6.8) {};
\node[draw, circle, minimum size=0.5cm, fill=wht] at (8.4, 6.8) {};
\node[draw, circle, minimum size=0.5cm, fill=wht] at (9.2, 6.8) {};
\node[draw, circle, minimum size=0.5cm, fill=wht] at (7.6, 3.6) {};
\node[draw, circle, minimum size=0.5cm, fill=wht] at (8.4, 3.6) {};
\node[draw, circle, minimum size=0.5cm, fill=wht] at (9.2, 3.6) {};
\node[draw, circle, minimum size=0.5cm, fill=wht] at (7.6, 2.8) {};
\node[draw, circle, minimum size=0.5cm, fill=wht] at (8.4, 2.8) {};
\node[draw, circle, minimum size=0.5cm, fill=wht] at (9.2, 2.8) {};
%----x4----
\node[draw, circle, minimum size=0.5cm, fill=wht] at (10.8, 3.6) {};
\node[draw, circle, minimum size=0.5cm, fill=wht] at (11.6, 3.6) {};
\node[draw, circle, minimum size=0.5cm, fill=wht] at (10.8, 2.8) {};
\node[draw, circle, minimum size=0.5cm, fill=wht] at (11.6, 2.8) {};
%--------G1--------
%----x1----
\node (end) at (0.9, 7.6) [right, font=\normalsize] {1};
\node (end) at (0.9, 6.8) [right, font=\normalsize] {1};
\node (end) at (0.9, 3.6) [right, font=\normalsize] {1};
\node (end) at (0.9, 2.8) [right, font=\normalsize] {1};
\node (end) at (2.5, 7.6) [right, font=\normalsize] {1};
\node (end) at (2.5, 6.8) [right, font=\normalsize] {1};
\node (end) at (2.5, 3.6) [right, font=\normalsize] {1};
\node (end) at (2.5, 2.8) [right, font=\normalsize] {1};
%----x2----
\node (end) at (4.1, 7.6) [right, font=\normalsize] {1};
\node (end) at (4.1, 6.8) [right, font=\normalsize] {1};
\node (end) at (5.7, 7.6) [right, font=\normalsize] {1};
\node (end) at (5.7, 6.8) [right, font=\normalsize] {1};
%----x3----
\node (end) at (7.3, 7.6) [right, font=\normalsize] {1};
\node (end) at (7.3, 6.8) [right, font=\normalsize] {1};
\node (end) at (7.3, 3.6) [right, font=\normalsize] {1};
\node (end) at (7.3, 2.8) [right, font=\normalsize] {1};
\node (end) at (8.9, 7.6) [right, font=\normalsize] {1};
\node (end) at (8.9, 6.8) [right, font=\normalsize] {1};
\node (end) at (8.9, 3.6) [right, font=\normalsize] {1};
\node (end) at (8.9, 2.8) [right, font=\normalsize] {1};
%----x4----
\node (end) at (10.5, 3.6) [right, font=\normalsize] {1};
\node (end) at (10.5, 2.8) [right, font=\normalsize] {1};

\draw[line width=0.6mm] (0, 0.8) -- (12.8, 0.8);
\draw[line width=0.6mm] (0, 2.4) -- (12.8, 2.4);
\draw[line width=0.6mm] (0, 4.8) -- (12.8, 4.8);
\draw[line width=0.6mm] (0, 6.4) -- (12.8, 6.4);

\draw[line width=0.6mm] (0, 8.8) -- (12.8, 8.8);
\draw[line width=0.6mm] (0, 10.4) -- (12.8, 10.4);

\draw[line width=0.6mm] (0, 0.8) -- (0, 2.4);
\draw[line width=0.6mm] (0, 4.8) -- (0, 6.4);
\draw[line width=0.6mm] (0, 8.8) -- (0, 10.4);

\draw[line width=0.6mm] (12.8, 0.8) -- (12.8, 2.4);
\draw[line width=0.6mm] (12.8, 4.8) -- (12.8, 6.4);
\draw[line width=0.6mm] (12.8, 8.8) -- (12.8, 10.4);
\end{tikzpicture}
\caption{Board after modifying G1}
\label{fig:G1}
\vspace{-5mm}
\end{figure}

Finally, in Figure~\ref{fig:G1}, 
we show the board obtained by applying the above modification for all the clauses.

% ---------------- Fig:G1 ----------------

\subsection{Arrangement of components in gadget G2} \label{sec:g2}
This gadget is for resolving P3. The board in Figure~\ref{fig:G1} rejects solutions in which three of the literal circles in each clause are all black or white, 
but accepts solutions in which one or two are black (i.e., one or two literals are true in 1-in-3SAT).
This means that 
a one-to-one correspondence fails to hold between the solution sets of Oredango and 1-in-3SAT.

To match the solutions of 1-in-3SAT and Oredango, we must accept only solutions 
where exactly one of the literal circles in the clause is black.
In order to resolve this problem, we place circles in gadget G2 and move
support circles from G1 to G2 appropriately.
Figure~\ref{fig:G2} shows
the proposed placement of circles in G1 and G2 for clause $C_1$.
Notice that the support circles at the coordinates (3,3) and (3,11) of G1 in Figure~\ref{fig:G1} 
are lifted to G2. 
This board achieves the desired property for the following reason.
In Figure~\ref{fig:G2},
given the colors of the three literal circles in the first line, 
the colors of any other circles are uniquely determined. 
Figs.~\ref{fig:G2-reject} and \ref{fig:G2-accept} show all possible coloring-patterns of Figure~\ref{fig:G2}
that are rejected and accepted, respectively. 
Notice that the coloring-patterns in 
Figs.~\ref{fig:G2-reject} and \ref{fig:G2-accept} can be interpreted as rejected and accepted solutions by 1-in-3SAT, respectively.
Also, notice that Figure~\ref{fig:G2-reject} covers any coloring-patterns such that two of the three literal circles in the first line are black.
This implies that P3 is resolved.
Hence, we have established a one-to-one correspondence between the two solution sets of Oredango and 1-in-3SAT.
Figure~\ref{fig:afterG2} shows the board after the above modification in G2.

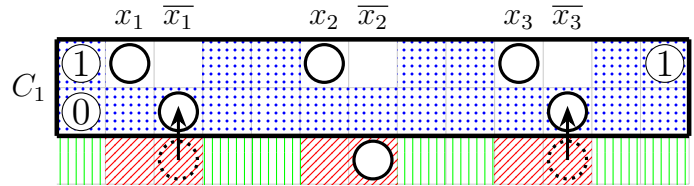
\begin{figure}[H]
\centering
\begin{tikzpicture}[scale=0.8]
%grid
\draw[step=0.8cm, color={rgb:black, 1;white, 4}] (-0.8, 0.0) grid (9.6, 2.4);
%Gadget
\path[pattern=north east lines, pattern color=red] (0.0, 0.0) rectangle + (1.6, 0.8);
\path[pattern=north east lines, pattern color=red] (3.2, 0.0) rectangle + (1.6, 0.8);
\path[pattern=north east lines, pattern color=red] (6.4, 0.0) rectangle + (1.6, 0.8);
\path[pattern=dots, pattern color=blue] (-0.8, 1.6) rectangle + (0.8, 0.8);
\path[pattern=dots, pattern color=blue] (1.6, 1.6) rectangle + (1.6, 0.8);
\path[pattern=dots, pattern color=blue] (4.8, 1.6) rectangle + (1.6, 0.8);
\path[pattern=dots, pattern color=blue] (8.0, 1.6) rectangle + (1.6, 0.8);
\path[pattern=dots, pattern color=blue] (-0.8, 0.8) rectangle + (10.4, 0.8);
\path[pattern=vertical lines, pattern color=green] (-0.8, 0.0) rectangle + (0.8, 0.8);
\path[pattern=vertical lines, pattern color=green] (1.6, 0.0) rectangle + (1.6, 0.8);
\path[pattern=vertical lines, pattern color=green] (4.8, 0.0) rectangle + (1.6, 0.8);
\path[pattern=vertical lines, pattern color=green] (8.0, 0.0) rectangle + (1.6, 0.8);
\draw[line width=0.6mm] (-0.8, 0.8) -- (9.6, 0.8);
\draw[line width=0.6mm] (-0.8, 2.4) -- (9.6, 2.4);
\draw[line width=0.6mm] (-0.8, 2.4) -- (-0.8, 0.8);
\draw[line width=0.6mm] (9.6, 2.4) -- (9.6, 0.8);
\node (end) at (-0.8, 1.6) [left] {\large $C_1$};
\node (end) at (0.4, 2.4) [above] {\large $x_{1}$};
\node (end) at (1.2, 2.4) [above] {\large $\overline{x_{1}}$};
\node (end) at (3.6, 2.4) [above] {\large $x_{2}$};
\node (end) at (4.4, 2.4) [above] {\large $\overline{x_{2}}$};
\node (end) at (6.8, 2.4) [above] {\large $x_{3}$};
\node (end) at (7.6, 2.4) [above] {\large $\overline{x_{3}}$};
\node[draw, very thick, circle, minimum size=0.5cm, fill=wht] at (0.4, 2.0) {};
\node[draw, very thick, circle, minimum size=0.5cm, fill=wht] at (1.2, 1.2) {};
\node[draw, very thick, dotted, circle, minimum size=0.5cm] at (1.2, 0.4) {};
\node[draw, circle, minimum size=0.5cm, fill=wht] at (-0.4, 2.0) {};
\node[draw, very thick, circle, minimum size=0.5cm, fill=wht] at (3.6, 2.0) {};
\node[draw, very thick, circle, minimum size=0.5cm, fill=wht] at (4.4, 0.4) {};
\node[draw, circle, minimum size=0.5cm, fill=wht] at (-0.4, 1.2) {};
\node[draw, very thick, circle, minimum size=0.5cm, fill=wht] at (6.8, 2.0) {};
\node[draw, very thick, circle, minimum size=0.5cm, fill=wht] at (7.6, 1.2) {};
\node[draw, very thick, dotted, circle, minimum size=0.5cm] at (7.6, 0.4) {};
\node[draw, circle, minimum size=0.5cm, fill=wht] at (9.2, 2.0) {};
\node[circle, black] at (-0.4, 2.0) {\Large $1$};
\node[circle, black] at (-0.4, 1.2) {\Large $0$};
\node[circle, black] at (9.2, 2.0) {\Large $1$};
\draw [line width=0.4mm, arrows = {-Stealth[scale=1.0]}]  (1.2,0.4) -- (1.2,1.3);
\draw [line width=0.4mm, arrows = {-Stealth[scale=1.0]}]  (7.6,0.4) -- (7.6,1.3);
\end{tikzpicture}
\caption{Placement of circles in G1 and G2 for $C_1 = \{x_1, x_2, x_3\}$}
\label{fig:G2}
\vspace{-5mm}
\end{figure}
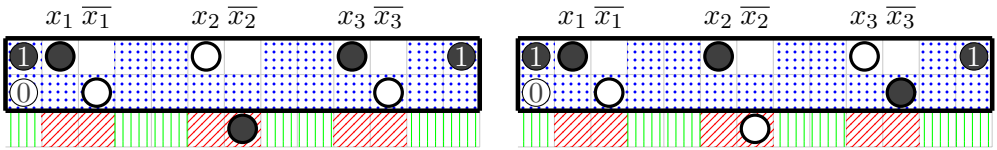
\begin{figure}[H]
\centering
\begin{tikzpicture}[scale=0.6]
\draw[step=0.8cm, color={rgb:black, 1;white, 4}] (-0.8, 0.0) grid (9.6, 2.4);
%Gadget
\path[pattern=north east lines, pattern color=red] (0.0, 0.0) rectangle + (1.6, 0.8);
\path[pattern=north east lines, pattern color=red] (3.2, 0.0) rectangle + (1.6, 0.8);
\path[pattern=north east lines, pattern color=red] (6.4, 0.0) rectangle + (1.6, 0.8);
\path[pattern=dots, pattern color=blue] (-0.8, 1.6) rectangle + (0.8, 0.8);
\path[pattern=dots, pattern color=blue] (1.6, 1.6) rectangle + (1.6, 0.8);
\path[pattern=dots, pattern color=blue] (4.8, 1.6) rectangle + (1.6, 0.8);
\path[pattern=dots, pattern color=blue] (8.0, 1.6) rectangle + (1.6, 0.8);
\path[pattern=dots, pattern color=blue] (-0.8, 0.8) rectangle + (10.4, 0.8);
\path[pattern=vertical lines, pattern color=green] (-0.8, 0.0) rectangle + (0.8, 0.8);
\path[pattern=vertical lines, pattern color=green] (1.6, 0.0) rectangle + (1.6, 0.8);
\path[pattern=vertical lines, pattern color=green] (4.8, 0.0) rectangle + (1.6, 0.8);
\path[pattern=vertical lines, pattern color=green] (8.0, 0.0) rectangle + (1.6, 0.8);
\draw[line width=0.6mm] (-0.8, 0.8) -- (9.6, 0.8);
\draw[line width=0.6mm] (-0.8, 2.4) -- (9.6, 2.4);
\draw[line width=0.6mm] (-0.8, 2.4) -- (-0.8, 0.8);
\draw[line width=0.6mm] (9.6, 2.4) -- (9.6, 0.8);
\node (end) at (0.4, 2.4) [above] {$x_{1}$};
\node (end) at (1.2, 2.4) [above] {$\overline{x_{1}}$};
\node (end) at (3.6, 2.4) [above] {$x_{2}$};
\node (end) at (4.4, 2.4) [above] {$\overline{x_{2}}$};
\node (end) at (6.8, 2.4) [above] {$x_{3}$};
\node (end) at (7.6, 2.4) [above] {$\overline{x_{3}}$};
\node[draw, very thick, circle, minimum size=0.3cm, fill=blk] at (0.4, 2.0) {};
\node[draw, very thick, circle, minimum size=0.3cm, fill=wht] at (1.2, 1.2) {};
\node[draw, circle, minimum size=0.3cm, fill=blk] at (-0.4, 2.0) {};
\node[draw, very thick, circle, minimum size=0.3cm, fill=wht] at (3.6, 2.0) {};
\node[draw, very thick, circle, minimum size=0.3cm, fill=blk] at (4.4, 0.4) {};
\node[draw, circle, minimum size=0.3cm, fill=wht] at (-0.4, 1.2) {};
\node[draw, very thick, circle, minimum size=0.3cm, fill=blk] at (6.8, 2.0) {};
\node[draw, very thick, circle, minimum size=0.3cm, fill=wht] at (7.6, 1.2) {};
\node[draw, circle, minimum size=0.3cm, fill=blk] at (9.2, 2.0) {};
\node[circle, wht] at (-0.4, 2.0) {$1$};
\node[circle, blk] at (-0.4, 1.2) {$0$};
\node[circle, wht] at (9.2, 2.0) {$1$};
\end{tikzpicture}
\hspace{0.05cm}
\begin{tikzpicture}[scale=0.6]
\draw[step=0.8cm, color={rgb:black, 1;white, 4}] (-0.8, 0.0) grid (9.6, 2.4);
%Gadget
\path[pattern=north east lines, pattern color=red] (0.0, 0.0) rectangle + (1.6, 0.8);
\path[pattern=north east lines, pattern color=red] (3.2, 0.0) rectangle + (1.6, 0.8);
\path[pattern=north east lines, pattern color=red] (6.4, 0.0) rectangle + (1.6, 0.8);
\path[pattern=dots, pattern color=blue] (-0.8, 1.6) rectangle + (0.8, 0.8);
\path[pattern=dots, pattern color=blue] (1.6, 1.6) rectangle + (1.6, 0.8);
\path[pattern=dots, pattern color=blue] (4.8, 1.6) rectangle + (1.6, 0.8);
\path[pattern=dots, pattern color=blue] (8.0, 1.6) rectangle + (1.6, 0.8);
\path[pattern=dots, pattern color=blue] (-0.8, 0.8) rectangle + (10.4, 0.8);
\path[pattern=vertical lines, pattern color=green] (-0.8, 0.0) rectangle + (0.8, 0.8);
\path[pattern=vertical lines, pattern color=green] (1.6, 0.0) rectangle + (1.6, 0.8);
\path[pattern=vertical lines, pattern color=green] (4.8, 0.0) rectangle + (1.6, 0.8);
\path[pattern=vertical lines, pattern color=green] (8.0, 0.0) rectangle + (1.6, 0.8);
\draw[line width=0.6mm] (-0.8, 0.8) -- (9.6, 0.8);
\draw[line width=0.6mm] (-0.8, 2.4) -- (9.6, 2.4);
\draw[line width=0.6mm] (-0.8, 2.4) -- (-0.8, 0.8);
\draw[line width=0.6mm] (9.6, 2.4) -- (9.6, 0.8);
\node (end) at (0.4, 2.4) [above] {$x_{1}$};
\node (end) at (1.2, 2.4) [above] {$\overline{x_{1}}$};
\node (end) at (3.6, 2.4) [above] {$x_{2}$};
\node (end) at (4.4, 2.4) [above] {$\overline{x_{2}}$};
\node (end) at (6.8, 2.4) [above] {$x_{3}$};
\node (end) at (7.6, 2.4) [above] {$\overline{x_{3}}$};
\node[draw, very thick, circle, minimum size=0.3cm, fill=blk] at (0.4, 2.0) {};
\node[draw, very thick, circle, minimum size=0.3cm, fill=wht] at (1.2, 1.2) {};
\node[draw, circle, minimum size=0.3cm, fill=blk] at (-0.4, 2.0) {};
\node[draw, very thick, circle, minimum size=0.3cm, fill=blk] at (3.6, 2.0) {};
\node[draw, very thick, circle, minimum size=0.3cm, fill=wht] at (4.4, 0.4) {};
\node[draw, circle, minimum size=0.3cm, fill=wht] at (-0.4, 1.2) {};
\node[draw, very thick, circle, minimum size=0.3cm, fill=wht] at (6.8, 2.0) {};
\node[draw, very thick, circle, minimum size=0.3cm, fill=blk] at (7.6, 1.2) {};
\node[draw, circle, minimum size=0.3cm, fill=blk] at (9.2, 2.0) {};
\node[circle, wht] at (-0.4, 2.0) {$1$};
\node[circle, blk] at (-0.4, 1.2) {$0$};
\node[circle, wht] at (9.2, 2.0) {$1$};
\end{tikzpicture}
\caption{Rejected patterns of Figure~\ref{fig:G2} (violating rule (c))}
\label{fig:G2-reject}
\vspace{-5mm}
\end{figure}
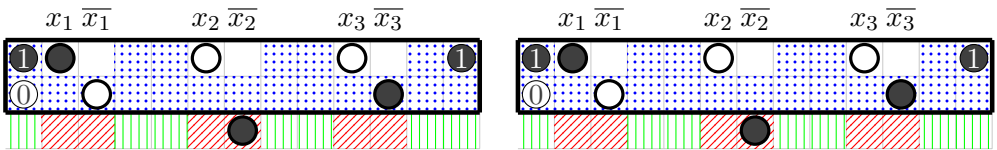
\begin{figure}[H]
\centering
\begin{tikzpicture}[scale=0.6]
\draw[step=0.8cm, color={rgb:black, 1;white, 4}] (-0.8, 0.0) grid (9.6, 2.4);
%Gadget
\path[pattern=north east lines, pattern color=red] (0.0, 0.0) rectangle + (1.6, 0.8);
\path[pattern=north east lines, pattern color=red] (3.2, 0.0) rectangle + (1.6, 0.8);
\path[pattern=north east lines, pattern color=red] (6.4, 0.0) rectangle + (1.6, 0.8);
\path[pattern=dots, pattern color=blue] (-0.8, 1.6) rectangle + (0.8, 0.8);
\path[pattern=dots, pattern color=blue] (1.6, 1.6) rectangle + (1.6, 0.8);
\path[pattern=dots, pattern color=blue] (4.8, 1.6) rectangle + (1.6, 0.8);
\path[pattern=dots, pattern color=blue] (8.0, 1.6) rectangle + (1.6, 0.8);
\path[pattern=dots, pattern color=blue] (-0.8, 0.8) rectangle + (10.4, 0.8);
\path[pattern=vertical lines, pattern color=green] (-0.8, 0.0) rectangle + (0.8, 0.8);
\path[pattern=vertical lines, pattern color=green] (1.6, 0.0) rectangle + (1.6, 0.8);
\path[pattern=vertical lines, pattern color=green] (4.8, 0.0) rectangle + (1.6, 0.8);
\path[pattern=vertical lines, pattern color=green] (8.0, 0.0) rectangle + (1.6, 0.8);
\draw[line width=0.6mm] (-0.8, 0.8) -- (9.6, 0.8);
\draw[line width=0.6mm] (-0.8, 2.4) -- (9.6, 2.4);
\draw[line width=0.6mm] (-0.8, 2.4) -- (-0.8, 0.8);
\draw[line width=0.6mm] (9.6, 2.4) -- (9.6, 0.8);
\node (end) at (0.4, 2.4) [above] {$x_{1}$};
\node (end) at (1.2, 2.4) [above] {$\overline{x_{1}}$};
\node (end) at (3.6, 2.4) [above] {$x_{2}$};
\node (end) at (4.4, 2.4) [above] {$\overline{x_{2}}$};
\node (end) at (6.8, 2.4) [above] {$x_{3}$};
\node (end) at (7.6, 2.4) [above] {$\overline{x_{3}}$};
\node[draw, very thick, circle, minimum size=0.3cm, fill=blk] at (0.4, 2.0) {};
\node[draw, very thick, circle, minimum size=0.3cm, fill=wht] at (1.2, 1.2) {};
\node[draw, circle, minimum size=0.3cm, fill=blk] at (-0.4, 2.0) {};
\node[draw, very thick, circle, minimum size=0.3cm, fill=wht] at (3.6, 2.0) {};
\node[draw, very thick, circle, minimum size=0.3cm, fill=blk] at (4.4, 0.4) {};
\node[draw, circle, minimum size=0.3cm, fill=wht] at (-0.4, 1.2) {};
\node[draw, very thick, circle, minimum size=0.3cm, fill=wht] at (6.8, 2.0) {};
\node[draw, very thick, circle, minimum size=0.3cm, fill=blk] at (7.6, 1.2) {};
\node[draw, circle, minimum size=0.3cm, fill=blk] at (9.2, 2.0) {};
\node[circle, wht] at (-0.4, 2.0) {$1$};
\node[circle, blk] at (-0.4, 1.2) {$0$};
\node[circle, wht] at (9.2, 2.0) {$1$};
\end{tikzpicture}
\hspace{0.05cm}
\begin{tikzpicture}[scale=0.6]
\draw[step=0.8cm, color={rgb:black, 1;white, 4}] (-0.8, 0.0) grid (9.6, 2.4);
%Gadget
\path[pattern=north east lines, pattern color=red] (0.0, 0.0) rectangle + (1.6, 0.8);
\path[pattern=north east lines, pattern color=red] (3.2, 0.0) rectangle + (1.6, 0.8);
\path[pattern=north east lines, pattern color=red] (6.4, 0.0) rectangle + (1.6, 0.8);
\path[pattern=dots, pattern color=blue] (-0.8, 1.6) rectangle + (0.8, 0.8);
\path[pattern=dots, pattern color=blue] (1.6, 1.6) rectangle + (1.6, 0.8);
\path[pattern=dots, pattern color=blue] (4.8, 1.6) rectangle + (1.6, 0.8);
\path[pattern=dots, pattern color=blue] (8.0, 1.6) rectangle + (1.6, 0.8);
\path[pattern=dots, pattern color=blue] (-0.8, 0.8) rectangle + (10.4, 0.8);
\path[pattern=vertical lines, pattern color=green] (-0.8, 0.0) rectangle + (0.8, 0.8);
\path[pattern=vertical lines, pattern color=green] (1.6, 0.0) rectangle + (1.6, 0.8);
\path[pattern=vertical lines, pattern color=green] (4.8, 0.0) rectangle + (1.6, 0.8);
\path[pattern=vertical lines, pattern color=green] (8.0, 0.0) rectangle + (1.6, 0.8);
\draw[line width=0.6mm] (-0.8, 0.8) -- (9.6, 0.8);
\draw[line width=0.6mm] (-0.8, 2.4) -- (9.6, 2.4);
\draw[line width=0.6mm] (-0.8, 2.4) -- (-0.8, 0.8);
\draw[line width=0.6mm] (9.6, 2.4) -- (9.6, 0.8);
\node (end) at (0.4, 2.4) [above] {$x_{1}$};
\node (end) at (1.2, 2.4) [above] {$\overline{x_{1}}$};
\node (end) at (3.6, 2.4) [above] {$x_{2}$};
\node (end) at (4.4, 2.4) [above] {$\overline{x_{2}}$};
\node (end) at (6.8, 2.4) [above] {$x_{3}$};
\node (end) at (7.6, 2.4) [above] {$\overline{x_{3}}$};
\node[draw, very thick, circle, minimum size=0.3cm, fill=blk] at (0.4, 2.0) {};
\node[draw, very thick, circle, minimum size=0.3cm, fill=wht] at (1.2, 1.2) {};
\node[draw, circle, minimum size=0.3cm, fill=blk] at (-0.4, 2.0) {};
\node[draw, very thick, circle, minimum size=0.3cm, fill=wht] at (3.6, 2.0) {};
\node[draw, very thick, circle, minimum size=0.3cm, fill=blk] at (4.4, 0.4) {};
\node[draw, circle, minimum size=0.3cm, fill=wht] at (-0.4, 1.2) {};
\node[draw, very thick, circle, minimum size=0.3cm, fill=wht] at (6.8, 2.0) {};
\node[draw, very thick, circle, minimum size=0.3cm, fill=blk] at (7.6, 1.2) {};
\node[draw, circle, minimum size=0.3cm, fill=blk] at (9.2, 2.0) {};
\node[circle, wht] at (-0.4, 2.0) {$1$};
\node[circle, blk] at (-0.4, 1.2) {$0$};
\node[circle, wht] at (9.2, 2.0) {$1$};
\end{tikzpicture}
\caption{Accepted patterns of Figure~\ref{fig:G2}}
\label{fig:G2-accept}
\vspace{-2mm}
\end{figure}

% ---------------- Fig:G2 ----------------
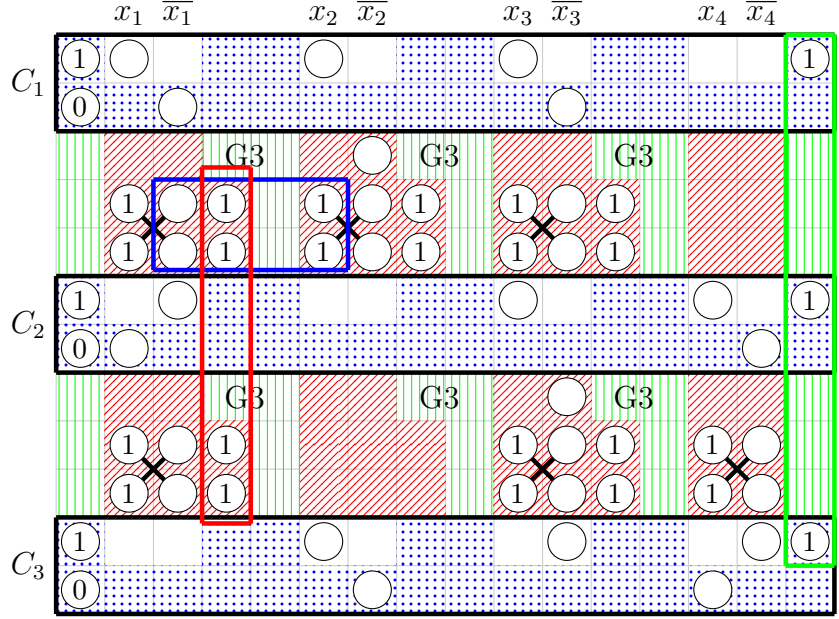
\begin{figure}[ht]
\centering
\begin{tikzpicture}[scale=0.8]

%grid
\draw[step=0.8cm, color={rgb:black, 1;white, 4}] (0, 0.8) grid (12.8, 10.4);

\node (end) at (0, 1.6) [left] {\large $C_{3}$};
\node (end) at (0, 5.6) [left] {\large $C_{2}$};
\node (end) at (0, 9.6) [left] {\large $C_{1}$};

\node (end) at (1.2, 10.4) [above] {\large $x_{1}$};
\node (end) at (2.0, 10.4) [above] {\large $\overline{x_{1}}$};
\node (end) at (4.4, 10.4) [above] {\large $x_{2}$};
\node (end) at (5.2, 10.4) [above] {\large $\overline{x_{2}}$};
\node (end) at (7.6, 10.4) [above] {\large $x_{3}$};
\node (end) at (8.4, 10.4) [above] {\large $\overline{x_{3}}$};
\node (end) at (10.8, 10.4) [above] {\large $x_{4}$};
\node (end) at (11.6, 10.4) [above] {\large $\overline{x_{4}}$};

%-----------------ガジェットG1-----------------%
\path[pattern=north east lines, pattern color=red] (0.8, 2.4) rectangle + (1.6, 2.4);
\path[pattern=north east lines, pattern color=red] (2.4, 2.4) rectangle + (0.8, 1.6);
\path[pattern=north east lines, pattern color=red] (0.8, 6.4) rectangle + (1.6, 2.4);
\path[pattern=north east lines, pattern color=red] (2.4, 6.4) rectangle + (0.8, 1.6);

\path[pattern=north east lines, pattern color=red] (4.0, 2.4) rectangle + (1.6, 2.4);
\path[pattern=north east lines, pattern color=red] (5.6, 2.4) rectangle + (0.8, 1.6);
\path[pattern=north east lines, pattern color=red] (4.0, 6.4) rectangle + (1.6, 2.4);
\path[pattern=north east lines, pattern color=red] (5.6, 6.4) rectangle + (0.8, 1.6);

\path[pattern=north east lines, pattern color=red] (7.2, 2.4) rectangle + (1.6, 2.4);
\path[pattern=north east lines, pattern color=red] (8.8, 2.4) rectangle + (0.8, 1.6);
\path[pattern=north east lines, pattern color=red] (7.2, 6.4) rectangle + (1.6, 2.4);
\path[pattern=north east lines, pattern color=red] (8.8, 6.4) rectangle + (0.8, 1.6);

\path[pattern=north east lines, pattern color=red] (10.4, 2.4) rectangle + (1.6, 2.4);
\path[pattern=north east lines, pattern color=red] (10.4, 6.4) rectangle + (1.6, 2.4);
%-----------------ガジェットG2-----------------%
\path[pattern=dots, pattern color=blue] (0.0, 1.6) rectangle + (0.8, 0.8);
\path[pattern=dots, pattern color=blue] (0.0, 5.6) rectangle + (0.8, 0.8);
\path[pattern=dots, pattern color=blue] (0.0, 9.6) rectangle + (0.8, 0.8);

\path[pattern=dots, pattern color=blue] (2.4, 1.6) rectangle + (1.6, 0.8);
\path[pattern=dots, pattern color=blue] (2.4, 5.6) rectangle + (1.6, 0.8);
\path[pattern=dots, pattern color=blue] (2.4, 9.6) rectangle + (1.6, 0.8);

\path[pattern=dots, pattern color=blue] (5.6, 1.6) rectangle + (1.6, 0.8);
\path[pattern=dots, pattern color=blue] (5.6, 5.6) rectangle + (1.6, 0.8);
\path[pattern=dots, pattern color=blue] (5.6, 9.6) rectangle + (1.6, 0.8);

\path[pattern=dots, pattern color=blue] (8.8, 1.6) rectangle + (1.6, 0.8);
\path[pattern=dots, pattern color=blue] (8.8, 5.6) rectangle + (1.6, 0.8);
\path[pattern=dots, pattern color=blue] (8.8, 9.6) rectangle + (1.6, 0.8);

\path[pattern=dots, pattern color=blue] (12.0, 1.6) rectangle + (0.8, 0.8);
\path[pattern=dots, pattern color=blue] (12.0, 5.6) rectangle + (0.8, 0.8);
\path[pattern=dots, pattern color=blue] (12.0, 9.6) rectangle + (0.8, 0.8);

\path[pattern=dots, pattern color=blue] (0.0, 0.8) rectangle + (12.8, 0.8);
\path[pattern=dots, pattern color=blue] (0.0, 4.8) rectangle + (12.8, 0.8);
\path[pattern=dots, pattern color=blue] (0.0, 8.8) rectangle + (12.8, 0.8);
%-----------------ガジェットG3-----------------%
\path[pattern=vertical lines, pattern color=green] (0.0, 2.4) rectangle + (0.8, 2.4);
\path[pattern=vertical lines, pattern color=green] (0.0, 6.4) rectangle + (0.8, 2.4);

\path[pattern=vertical lines, pattern color=green] (3.2, 2.4) rectangle + (0.8, 1.6);
\path[pattern=vertical lines, pattern color=green] (2.4, 4.0) rectangle + (1.6, 0.8);
\path[pattern=vertical lines, pattern color=green] (3.2, 6.4) rectangle + (0.8, 1.6);
\path[pattern=vertical lines, pattern color=green] (2.4, 8.0) rectangle + (1.6, 0.8);

\path[pattern=vertical lines, pattern color=green] (6.4, 2.4) rectangle + (0.8, 1.6);
\path[pattern=vertical lines, pattern color=green] (5.6, 4.0) rectangle + (1.6, 0.8);
\path[pattern=vertical lines, pattern color=green] (6.4, 6.4) rectangle + (0.8, 1.6);
\path[pattern=vertical lines, pattern color=green] (5.6, 8.0) rectangle + (1.6, 0.8);

\path[pattern=vertical lines, pattern color=green] (9.6, 2.4) rectangle + (0.8, 1.6);
\path[pattern=vertical lines, pattern color=green] (8.8, 4.0) rectangle + (1.6, 0.8);
\path[pattern=vertical lines, pattern color=green] (9.6, 6.4) rectangle + (0.8, 1.6);
\path[pattern=vertical lines, pattern color=green] (8.8, 8.0) rectangle + (1.6, 0.8);

\path[pattern=vertical lines, pattern color=green] (12.0, 2.4) rectangle + (0.8, 2.4);
\path[pattern=vertical lines, pattern color=green] (12.0, 6.4) rectangle + (0.8, 2.4);
%-----------串の描画-----------
\draw[line width=0.6mm] (1.2, 6.8) -- (2.0, 7.6);
\draw[line width=0.6mm] (1.2, 7.6) -- (2.0, 6.8);
\draw[line width=0.6mm] (1.2, 2.8) -- (2.0, 3.6);
\draw[line width=0.6mm] (1.2, 3.6) -- (2.0, 2.8);

\draw[line width=0.6mm] (4.4, 6.8) -- (5.2, 7.6);
\draw[line width=0.6mm] (4.4, 7.6) -- (5.2, 6.8);

\draw[line width=0.6mm] (7.6, 6.8) -- (8.4, 7.6);
\draw[line width=0.6mm] (7.6, 7.6) -- (8.4, 6.8);
\draw[line width=0.6mm] (7.6, 2.8) -- (8.4, 3.6);
\draw[line width=0.6mm] (7.6, 3.6) -- (8.4, 2.8);

\draw[line width=0.6mm] (10.8, 2.8) -- (11.6, 3.6);
\draw[line width=0.6mm] (10.8, 3.6) -- (11.6, 2.8);
%-----------------ガジェットG3-----------------%
\node (end) at (2.6, 4.4) [right, font=\large] {G3};
\node (end) at (2.6, 8.4) [right, font=\large] {G3};

\node (end) at (5.8, 4.4) [right, font=\large] {G3};
\node (end) at (5.8, 8.4) [right, font=\large] {G3};

\node (end) at (9.0, 4.4) [right, font=\large] {G3};
\node (end) at (9.0, 8.4) [right, font=\large] {G3};
%-----------だんごの描画-----------
%C_1
\node[draw, circle, minimum size=0.5cm, fill=wht] at (1.2, 10.0) {};
\node[draw, circle, minimum size=0.5cm, fill=wht] at (2.0, 9.2) {};
\node[draw, circle, minimum size=0.5cm, fill=wht] at (4.4, 10.0) {};
\node[draw, circle, minimum size=0.5cm, fill=wht] at (5.2, 8.4) {};
\node[draw, circle, minimum size=0.5cm, fill=wht] at (0.4, 9.2) {};
\node[draw, circle, minimum size=0.5cm, fill=wht] at (7.6, 10.0) {};
\node[draw, circle, minimum size=0.5cm, fill=wht] at (8.4, 9.2) {};
%C2
\node[draw, circle, minimum size=0.5cm, fill=wht] at (1.2, 5.2) {};
\node[draw, circle, minimum size=0.5cm, fill=wht] at (2.0, 6.0) {};
\node[draw, circle, minimum size=0.5cm, fill=wht] at (7.6, 6.0) {};
\node[draw, circle, minimum size=0.5cm, fill=wht] at (8.4, 4.4) {};
\node[draw, circle, minimum size=0.5cm, fill=wht] at (0.4, 5.2) {};
\node[draw, circle, minimum size=0.5cm, fill=wht] at (10.8, 6.0) {};
\node[draw, circle, minimum size=0.5cm, fill=wht] at (11.6, 5.2) {};
%C3
\node[draw, circle, minimum size=0.5cm, fill=wht] at (4.4, 2.0) {};
\node[draw, circle, minimum size=0.5cm, fill=wht] at (5.2, 1.2) {};
\node[draw, circle, minimum size=0.5cm, fill=wht] at (8.4, 2.0) {};
\node[draw, circle, minimum size=0.5cm, fill=wht] at (0.4, 1.2) {};
\node[draw, circle, minimum size=0.5cm, fill=wht] at (10.8, 1.2) {};
\node[draw, circle, minimum size=0.5cm, fill=wht] at (11.6, 2.0) {};
%--------G1--------
%----x1----
\node[draw, circle, minimum size=0.5cm, fill=wht] at (1.2, 7.6) {};
\node[draw, circle, minimum size=0.5cm, fill=wht] at (2.0, 7.6) {};
\node[draw, circle, minimum size=0.5cm, fill=wht] at (2.8, 7.6) {};
\node[draw, circle, minimum size=0.5cm, fill=wht] at (1.2, 6.8) {};
\node[draw, circle, minimum size=0.5cm, fill=wht] at (2.0, 6.8) {};
\node[draw, circle, minimum size=0.5cm, fill=wht] at (2.8, 6.8) {};
\node[draw, circle, minimum size=0.5cm, fill=wht] at (1.2, 3.6) {};
\node[draw, circle, minimum size=0.5cm, fill=wht] at (2.0, 3.6) {};
\node[draw, circle, minimum size=0.5cm, fill=wht] at (2.8, 3.6) {};
\node[draw, circle, minimum size=0.5cm, fill=wht] at (1.2, 2.8) {};
\node[draw, circle, minimum size=0.5cm, fill=wht] at (2.0, 2.8) {};
\node[draw, circle, minimum size=0.5cm, fill=wht] at (2.8, 2.8) {};
%----x2----
\node[draw, circle, minimum size=0.5cm, fill=wht] at (4.4, 7.6) {};
\node[draw, circle, minimum size=0.5cm, fill=wht] at (5.2, 7.6) {};
\node[draw, circle, minimum size=0.5cm, fill=wht] at (6.0, 7.6) {};
\node[draw, circle, minimum size=0.5cm, fill=wht] at (4.4, 6.8) {};
\node[draw, circle, minimum size=0.5cm, fill=wht] at (5.2, 6.8) {};
\node[draw, circle, minimum size=0.5cm, fill=wht] at (6.0, 6.8) {};
%----x3----
\node[draw, circle, minimum size=0.5cm, fill=wht] at (7.6, 7.6) {};
\node[draw, circle, minimum size=0.5cm, fill=wht] at (8.4, 7.6) {};
\node[draw, circle, minimum size=0.5cm, fill=wht] at (9.2, 7.6) {};
\node[draw, circle, minimum size=0.5cm, fill=wht] at (7.6, 6.8) {};
\node[draw, circle, minimum size=0.5cm, fill=wht] at (8.4, 6.8) {};
\node[draw, circle, minimum size=0.5cm, fill=wht] at (9.2, 6.8) {};
\node[draw, circle, minimum size=0.5cm, fill=wht] at (7.6, 3.6) {};
\node[draw, circle, minimum size=0.5cm, fill=wht] at (8.4, 3.6) {};
\node[draw, circle, minimum size=0.5cm, fill=wht] at (9.2, 3.6) {};
\node[draw, circle, minimum size=0.5cm, fill=wht] at (7.6, 2.8) {};
\node[draw, circle, minimum size=0.5cm, fill=wht] at (8.4, 2.8) {};
\node[draw, circle, minimum size=0.5cm, fill=wht] at (9.2, 2.8) {};
%----x4----
\node[draw, circle, minimum size=0.5cm, fill=wht] at (10.8, 3.6) {};
\node[draw, circle, minimum size=0.5cm, fill=wht] at (11.6, 3.6) {};
\node[draw, circle, minimum size=0.5cm, fill=wht] at (10.8, 2.8) {};
\node[draw, circle, minimum size=0.5cm, fill=wht] at (11.6, 2.8) {};
%--------G2--------
\node[draw, circle, minimum size=0.5cm, fill=wht] at (0.4, 10.0) {};
\node[draw, circle, minimum size=0.5cm, fill=wht] at (12.4, 10.0) {};
\node[draw, circle, minimum size=0.5cm, fill=wht] at (0.4, 6.0) {};
\node[draw, circle, minimum size=0.5cm, fill=wht] at (12.4, 6.0) {};
\node[draw, circle, minimum size=0.5cm, fill=wht] at (0.4, 2.0) {};
\node[draw, circle, minimum size=0.5cm, fill=wht] at (12.4, 2.0) {};
%--------G1--------
%----x1----
\node (end) at (0.9, 7.6) [right, font=\normalsize] {1};
\node (end) at (0.9, 6.8) [right, font=\normalsize] {1};
\node (end) at (0.9, 3.6) [right, font=\normalsize] {1};
\node (end) at (0.9, 2.8) [right, font=\normalsize] {1};
\node (end) at (2.5, 7.6) [right, font=\normalsize] {1};
\node (end) at (2.5, 6.8) [right, font=\normalsize] {1};
\node (end) at (2.5, 3.6) [right, font=\normalsize] {1};
\node (end) at (2.5, 2.8) [right, font=\normalsize] {1};
%----x2----
\node (end) at (4.1, 7.6) [right, font=\normalsize] {1};
\node (end) at (4.1, 6.8) [right, font=\normalsize] {1};
\node (end) at (5.7, 7.6) [right, font=\normalsize] {1};
\node (end) at (5.7, 6.8) [right, font=\normalsize] {1};
%----x3----
\node (end) at (7.3, 7.6) [right, font=\normalsize] {1};
\node (end) at (7.3, 6.8) [right, font=\normalsize] {1};
\node (end) at (7.3, 3.6) [right, font=\normalsize] {1};
\node (end) at (7.3, 2.8) [right, font=\normalsize] {1};
\node (end) at (8.9, 7.6) [right, font=\normalsize] {1};
\node (end) at (8.9, 6.8) [right, font=\normalsize] {1};
\node (end) at (8.9, 3.6) [right, font=\normalsize] {1};
\node (end) at (8.9, 2.8) [right, font=\normalsize] {1};
%----x4----
\node (end) at (10.5, 3.6) [right, font=\normalsize] {1};
\node (end) at (10.5, 2.8) [right, font=\normalsize] {1};
%--------G2--------
\node (end) at (0.1, 2.0) [right, font=\normalsize] {1};
\node (end) at (12.1, 2.0) [right, font=\normalsize] {1};
\node (end) at (0.1, 6.0) [right, font=\normalsize] {1};
\node (end) at (12.1, 6.0) [right, font=\normalsize] {1};
\node (end) at (0.1, 10.0) [right, font=\normalsize] {1};
\node (end) at (12.1, 10.0) [right, font=\normalsize] {1};

\node (end) at (0.1, 1.2) [right, font=\normalsize] {0};
\node (end) at (0.1, 5.2) [right, font=\normalsize] {0};
\node (end) at (0.1, 9.2) [right, font=\normalsize] {0};

\draw[line width=0.6mm] (0, 0.8) -- (12.8, 0.8);
\draw[line width=0.6mm] (0, 2.4) -- (12.8, 2.4);
\draw[line width=0.6mm] (0, 4.8) -- (12.8, 4.8);
\draw[line width=0.6mm] (0, 6.4) -- (12.8, 6.4);

\draw[line width=0.6mm] (0, 8.8) -- (12.8, 8.8);
\draw[line width=0.6mm] (0, 10.4) -- (12.8, 10.4);

\draw[line width=0.6mm] (0, 0.8) -- (0, 2.4);
\draw[line width=0.6mm] (0, 4.8) -- (0, 6.4);
\draw[line width=0.6mm] (0, 8.8) -- (0, 10.4);

\draw[line width=0.6mm] (12.8, 0.8) -- (12.8, 2.4);
\draw[line width=0.6mm] (12.8, 4.8) -- (12.8, 6.4);
\draw[line width=0.6mm] (12.8, 8.8) -- (12.8, 10.4);

\draw[line width=0.6mm, color=blue] (1.6, 6.5) -- (1.6, 8.0);
\draw[line width=0.6mm, color=blue] (4.8, 6.5) -- (4.8, 8.0);
\draw[line width=0.6mm, color=blue] (1.6, 6.5) -- (4.8, 6.5);
\draw[line width=0.6mm, color=blue] (1.6, 8.0) -- (4.8, 8.0);

\draw[line width=0.6mm, color=green] (12.0, 1.6) -- (12.8, 1.6);
\draw[line width=0.6mm, color=green] (12.8, 1.6) -- (12.8, 10.4);
\draw[line width=0.6mm, color=green] (12.0, 10.4) -- (12.8, 10.4);
\draw[line width=0.6mm, color=green] (12.0, 1.6) -- (12.0, 10.4);

\draw[line width=0.6mm, color=red] (2.4, 8.2) -- (2.4, 2.3);
\draw[line width=0.6mm, color=red] (3.2, 8.2) -- (3.2, 2.3);
\draw[line width=0.6mm, color=red] (2.4, 8.2) -- (3.2, 8.2);
\draw[line width=0.6mm, color=red] (2.4, 2.3) -- (3.2, 2.3);

\end{tikzpicture}
\caption{Board after modifying G2}
\label{fig:afterG2}
\vspace{-8.5mm}
\end{figure}

\subsection{Arrangement of components in gadget G3} \label{sec:g3}

\begin{figure}[ht]
\centering
\begin{tikzpicture}[scale=0.8]

%grid
\draw[step=0.8cm, color={rgb:black, 1;white, 4}] (0, 0.8) grid (12.8, 10.4);

\node (end) at (0, 1.6) [left] {$C_{3}$};
\node (end) at (0, 5.6) [left] {$C_{2}$};
\node (end) at (0, 9.6) [left] {$C_{1}$};

\node (end) at (1.2, 10.4) [above] {$x_{1}$};
\node (end) at (2.0, 10.4) [above] {$\overline{x_{1}}$};
\node (end) at (4.4, 10.4) [above] {$x_{2}$};
\node (end) at (5.2, 10.4) [above] {$\overline{x_{2}}$};
\node (end) at (7.6, 10.4) [above] {$x_{3}$};
\node (end) at (8.4, 10.4) [above] {$\overline{x_{3}}$};
\node (end) at (10.8, 10.4) [above] {$x_{4}$};
\node (end) at (11.6, 10.4) [above] {$\overline{x_{4}}$};

%-----------------ガジェットG1-----------------%
\path[pattern=north east lines, pattern color=red] (0.8, 2.4) rectangle + (1.6, 2.4);
\path[pattern=north east lines, pattern color=red] (2.4, 2.4) rectangle + (0.8, 1.6);
\path[pattern=north east lines, pattern color=red] (0.8, 6.4) rectangle + (1.6, 2.4);
\path[pattern=north east lines, pattern color=red] (2.4, 6.4) rectangle + (0.8, 1.6);

\path[pattern=north east lines, pattern color=red] (4.0, 2.4) rectangle + (1.6, 2.4);
\path[pattern=north east lines, pattern color=red] (5.6, 2.4) rectangle + (0.8, 1.6);
\path[pattern=north east lines, pattern color=red] (4.0, 6.4) rectangle + (1.6, 2.4);
\path[pattern=north east lines, pattern color=red] (5.6, 6.4) rectangle + (0.8, 1.6);

\path[pattern=north east lines, pattern color=red] (7.2, 2.4) rectangle + (1.6, 2.4);
\path[pattern=north east lines, pattern color=red] (8.8, 2.4) rectangle + (0.8, 1.6);
\path[pattern=north east lines, pattern color=red] (7.2, 6.4) rectangle + (1.6, 2.4);
\path[pattern=north east lines, pattern color=red] (8.8, 6.4) rectangle + (0.8, 1.6);

\path[pattern=north east lines, pattern color=red] (10.4, 2.4) rectangle + (1.6, 2.4);
\path[pattern=north east lines, pattern color=red] (10.4, 6.4) rectangle + (1.6, 2.4);
%-----------------ガジェットG2-----------------%
\path[pattern=dots, pattern color=blue] (0.0, 1.6) rectangle + (0.8, 0.8);
\path[pattern=dots, pattern color=blue] (0.0, 5.6) rectangle + (0.8, 0.8);
\path[pattern=dots, pattern color=blue] (0.0, 9.6) rectangle + (0.8, 0.8);

\path[pattern=dots, pattern color=blue] (2.4, 1.6) rectangle + (1.6, 0.8);
\path[pattern=dots, pattern color=blue] (2.4, 5.6) rectangle + (1.6, 0.8);
\path[pattern=dots, pattern color=blue] (2.4, 9.6) rectangle + (1.6, 0.8);

\path[pattern=dots, pattern color=blue] (5.6, 1.6) rectangle + (1.6, 0.8);
\path[pattern=dots, pattern color=blue] (5.6, 5.6) rectangle + (1.6, 0.8);
\path[pattern=dots, pattern color=blue] (5.6, 9.6) rectangle + (1.6, 0.8);

\path[pattern=dots, pattern color=blue] (8.8, 1.6) rectangle + (1.6, 0.8);
\path[pattern=dots, pattern color=blue] (8.8, 5.6) rectangle + (1.6, 0.8);
\path[pattern=dots, pattern color=blue] (8.8, 9.6) rectangle + (1.6, 0.8);

\path[pattern=dots, pattern color=blue] (12.0, 1.6) rectangle + (0.8, 0.8);
\path[pattern=dots, pattern color=blue] (12.0, 5.6) rectangle + (0.8, 0.8);
\path[pattern=dots, pattern color=blue] (12.0, 9.6) rectangle + (0.8, 0.8);

\path[pattern=dots, pattern color=blue] (0.0, 0.8) rectangle + (12.8, 0.8);
\path[pattern=dots, pattern color=blue] (0.0, 4.8) rectangle + (12.8, 0.8);
\path[pattern=dots, pattern color=blue] (0.0, 8.8) rectangle + (12.8, 0.8);
%-----------------ガジェットG3-----------------%
\path[pattern=vertical lines, pattern color=green] (0.0, 2.4) rectangle + (0.8, 2.4);
\path[pattern=vertical lines, pattern color=green] (0.0, 6.4) rectangle + (0.8, 2.4);

\path[pattern=vertical lines, pattern color=green] (3.2, 2.4) rectangle + (0.8, 1.6);
\path[pattern=vertical lines, pattern color=green] (2.4, 4.0) rectangle + (1.6, 0.8);
\path[pattern=vertical lines, pattern color=green] (3.2, 6.4) rectangle + (0.8, 1.6);
\path[pattern=vertical lines, pattern color=green] (2.4, 8.0) rectangle + (1.6, 0.8);

\path[pattern=vertical lines, pattern color=green] (6.4, 2.4) rectangle + (0.8, 1.6);
\path[pattern=vertical lines, pattern color=green] (5.6, 4.0) rectangle + (1.6, 0.8);
\path[pattern=vertical lines, pattern color=green] (6.4, 6.4) rectangle + (0.8, 1.6);
\path[pattern=vertical lines, pattern color=green] (5.6, 8.0) rectangle + (1.6, 0.8);

\path[pattern=vertical lines, pattern color=green] (9.6, 2.4) rectangle + (0.8, 1.6);
\path[pattern=vertical lines, pattern color=green] (8.8, 4.0) rectangle + (1.6, 0.8);
\path[pattern=vertical lines, pattern color=green] (9.6, 6.4) rectangle + (0.8, 1.6);
\path[pattern=vertical lines, pattern color=green] (8.8, 8.0) rectangle + (1.6, 0.8);

\path[pattern=vertical lines, pattern color=green] (12.0, 2.4) rectangle + (0.8, 2.4);
\path[pattern=vertical lines, pattern color=green] (12.0, 6.4) rectangle + (0.8, 2.4);
%-----------串の描画-----------
\draw[line width=0.6mm] (1.2, 6.8) -- (2.0, 7.6);
\draw[line width=0.6mm] (1.2, 7.6) -- (2.0, 6.8);
\draw[line width=0.6mm] (1.2, 2.8) -- (2.0, 3.6);
\draw[line width=0.6mm] (1.2, 3.6) -- (2.0, 2.8);

\draw[line width=0.6mm] (4.4, 6.8) -- (5.2, 7.6);
\draw[line width=0.6mm] (4.4, 7.6) -- (5.2, 6.8);

\draw[line width=0.6mm] (7.6, 6.8) -- (8.4, 7.6);
\draw[line width=0.6mm] (7.6, 7.6) -- (8.4, 6.8);
\draw[line width=0.6mm] (7.6, 2.8) -- (8.4, 3.6);
\draw[line width=0.6mm] (7.6, 3.6) -- (8.4, 2.8);

\draw[line width=0.6mm] (10.8, 2.8) -- (11.6, 3.6);
\draw[line width=0.6mm] (10.8, 3.6) -- (11.6, 2.8);
%-----------だんごの描画-----------
%C_1
\node[draw, circle, minimum size=0.5cm, fill=wht] at (1.2, 10.0) {};
\node[draw, circle, minimum size=0.5cm, fill=wht] at (2.0, 9.2) {};
\node[draw, circle, minimum size=0.5cm, fill=wht] at (4.4, 10.0) {};
\node[draw, circle, minimum size=0.5cm, fill=wht] at (5.2, 8.4) {};
\node[draw, circle, minimum size=0.5cm, fill=wht] at (0.4, 9.2) {};
\node[draw, circle, minimum size=0.5cm, fill=wht] at (7.6, 10.0) {};
\node[draw, circle, minimum size=0.5cm, fill=wht] at (8.4, 9.2) {};
%C2
\node[draw, circle, minimum size=0.5cm, fill=wht] at (1.2, 5.2) {};
\node[draw, circle, minimum size=0.5cm, fill=wht] at (2.0, 6.0) {};
\node[draw, circle, minimum size=0.5cm, fill=wht] at (7.6, 6.0) {};
\node[draw, circle, minimum size=0.5cm, fill=wht] at (8.4, 4.4) {};
\node[draw, circle, minimum size=0.5cm, fill=wht] at (0.4, 5.2) {};
\node[draw, circle, minimum size=0.5cm, fill=wht] at (10.8, 6.0) {};
\node[draw, circle, minimum size=0.5cm, fill=wht] at (11.6, 5.2) {};
%C3
\node[draw, circle, minimum size=0.5cm, fill=wht] at (4.4, 2.0) {};
\node[draw, circle, minimum size=0.5cm, fill=wht] at (5.2, 1.2) {};
\node[draw, circle, minimum size=0.5cm, fill=wht] at (8.4, 2.0) {};
\node[draw, circle, minimum size=0.5cm, fill=wht] at (0.4, 1.2) {};
\node[draw, circle, minimum size=0.5cm, fill=wht] at (10.8, 1.2) {};
\node[draw, circle, minimum size=0.5cm, fill=wht] at (11.6, 2.0) {};
%--------G1--------
%----x1----
\node[draw, circle, minimum size=0.5cm, fill=wht] at (1.2, 7.6) {};
\node[draw, circle, minimum size=0.5cm, fill=wht] at (2.0, 7.6) {};
\node[draw, circle, minimum size=0.5cm, fill=wht] at (2.8, 7.6) {};
\node[draw, circle, minimum size=0.5cm, fill=wht] at (1.2, 6.8) {};
\node[draw, circle, minimum size=0.5cm, fill=wht] at (2.0, 6.8) {};
\node[draw, circle, minimum size=0.5cm, fill=wht] at (2.8, 6.8) {};
\node[draw, circle, minimum size=0.5cm, fill=wht] at (1.2, 3.6) {};
\node[draw, circle, minimum size=0.5cm, fill=wht] at (2.0, 3.6) {};
\node[draw, circle, minimum size=0.5cm, fill=wht] at (2.8, 3.6) {};
\node[draw, circle, minimum size=0.5cm, fill=wht] at (1.2, 2.8) {};
\node[draw, circle, minimum size=0.5cm, fill=wht] at (2.0, 2.8) {};
\node[draw, circle, minimum size=0.5cm, fill=wht] at (2.8, 2.8) {};
%----x2----
\node[draw, circle, minimum size=0.5cm, fill=wht] at (4.4, 7.6) {};
\node[draw, circle, minimum size=0.5cm, fill=wht] at (5.2, 7.6) {};
\node[draw, circle, minimum size=0.5cm, fill=wht] at (6.0, 7.6) {};
\node[draw, circle, minimum size=0.5cm, fill=wht] at (4.4, 6.8) {};
\node[draw, circle, minimum size=0.5cm, fill=wht] at (5.2, 6.8) {};
\node[draw, circle, minimum size=0.5cm, fill=wht] at (6.0, 6.8) {};
%----x3----
\node[draw, circle, minimum size=0.5cm, fill=wht] at (7.6, 7.6) {};
\node[draw, circle, minimum size=0.5cm, fill=wht] at (8.4, 7.6) {};
\node[draw, circle, minimum size=0.5cm, fill=wht] at (9.2, 7.6) {};
\node[draw, circle, minimum size=0.5cm, fill=wht] at (7.6, 6.8) {};
\node[draw, circle, minimum size=0.5cm, fill=wht] at (8.4, 6.8) {};
\node[draw, circle, minimum size=0.5cm, fill=wht] at (9.2, 6.8) {};
\node[draw, circle, minimum size=0.5cm, fill=wht] at (7.6, 3.6) {};
\node[draw, circle, minimum size=0.5cm, fill=wht] at (8.4, 3.6) {};
\node[draw, circle, minimum size=0.5cm, fill=wht] at (9.2, 3.6) {};
\node[draw, circle, minimum size=0.5cm, fill=wht] at (7.6, 2.8) {};
\node[draw, circle, minimum size=0.5cm, fill=wht] at (8.4, 2.8) {};
\node[draw, circle, minimum size=0.5cm, fill=wht] at (9.2, 2.8) {};
%----x4----
\node[draw, circle, minimum size=0.5cm, fill=wht] at (10.8, 3.6) {};
\node[draw, circle, minimum size=0.5cm, fill=wht] at (11.6, 3.6) {};
\node[draw, circle, minimum size=0.5cm, fill=wht] at (10.8, 2.8) {};
\node[draw, circle, minimum size=0.5cm, fill=wht] at (11.6, 2.8) {};
%--------G2--------
\node[draw, circle, minimum size=0.5cm, fill=wht] at (0.4, 10.0) {};
\node[draw, circle, minimum size=0.5cm, fill=wht] at (12.4, 10.0) {};
\node[draw, circle, minimum size=0.5cm, fill=wht] at (0.4, 6.0) {};
\node[draw, circle, minimum size=0.5cm, fill=wht] at (12.4, 6.0) {};
\node[draw, circle, minimum size=0.5cm, fill=wht] at (0.4, 2.0) {};
\node[draw, circle, minimum size=0.5cm, fill=wht] at (12.4, 2.0) {};
%--------G3--------
%左端

%----x1----
\node[draw, circle, minimum size=0.5cm, fill=wht] at (2.8, 8.4) {};
\node[draw, circle, minimum size=0.5cm, fill=wht] at (3.6, 8.4) {};
\node[draw, circle, minimum size=0.5cm, fill=wht] at (3.6, 7.6) {};
\node[draw, circle, minimum size=0.5cm, fill=wht] at (3.6, 6.8) {};

\node[draw, circle, minimum size=0.5cm, fill=wht] at (2.8, 4.4) {};
\node[draw, circle, minimum size=0.5cm, fill=wht] at (3.6, 4.4) {};
\node[draw, circle, minimum size=0.5cm, fill=wht] at (3.6, 3.6) {};
\node[draw, circle, minimum size=0.5cm, fill=wht] at (3.6, 2.8) {};
%----x2----
\node[draw, circle, minimum size=0.5cm, fill=wht] at (6.0, 8.4) {};
\node[draw, circle, minimum size=0.5cm, fill=wht] at (6.8, 8.4) {};
\node[draw, circle, minimum size=0.5cm, fill=wht] at (6.8, 7.6) {};
\node[draw, circle, minimum size=0.5cm, fill=wht] at (6.8, 6.8) {};
%----x3----
\node[draw, circle, minimum size=0.5cm, fill=wht] at (9.2, 8.4) {};
\node[draw, circle, minimum size=0.5cm, fill=wht] at (10.0, 8.4) {};
\node[draw, circle, minimum size=0.5cm, fill=wht] at (10.0, 7.6) {};
\node[draw, circle, minimum size=0.5cm, fill=wht] at (10.0, 6.8) {};

\node[draw, circle, minimum size=0.5cm, fill=wht] at (9.2, 4.4) {};
\node[draw, circle, minimum size=0.5cm, fill=wht] at (10.0, 4.4) {};
\node[draw, circle, minimum size=0.5cm, fill=wht] at (10.0, 3.6) {};
\node[draw, circle, minimum size=0.5cm, fill=wht] at (10.0, 2.8) {};

\node[draw, circle, minimum size=0.5cm, fill=wht] at (12.4, 7.6) {};
\node[draw, circle, minimum size=0.5cm, fill=wht] at (12.4, 6.8) {};
\node[draw, circle, minimum size=0.5cm, fill=wht] at (12.4, 3.6) {};
\node[draw, circle, minimum size=0.5cm, fill=wht] at (12.4, 2.8) {};
%--------G1--------
%----x1----
\node (end) at (0.9, 7.6) [right, font=\normalsize] {1};
\node (end) at (0.9, 6.8) [right, font=\normalsize] {1};
\node (end) at (0.9, 3.6) [right, font=\normalsize] {1};
\node (end) at (0.9, 2.8) [right, font=\normalsize] {1};
%----x2----
\node (end) at (4.1, 7.6) [right, font=\normalsize] {1};
\node (end) at (4.1, 6.8) [right, font=\normalsize] {1};
%----x3----
\node (end) at (7.3, 7.6) [right, font=\normalsize] {1};
\node (end) at (7.3, 6.8) [right, font=\normalsize] {1};
\node (end) at (7.3, 3.6) [right, font=\normalsize] {1};
\node (end) at (7.3, 2.8) [right, font=\normalsize] {1};
%----x4----
\node (end) at (10.5, 3.6) [right, font=\normalsize] {1};
\node (end) at (10.5, 2.8) [right, font=\normalsize] {1};
%--------G2--------
\node (end) at (0.1, 2.0) [right, font=\normalsize] {1};
\node (end) at (12.1, 2.0) [right, font=\normalsize] {1};
\node (end) at (0.1, 6.0) [right, font=\normalsize] {1};
\node (end) at (12.1, 6.0) [right, font=\normalsize] {1};
\node (end) at (0.1, 10.0) [right, font=\normalsize] {1};
\node (end) at (12.1, 10.0) [right, font=\normalsize] {1};

\node (end) at (0.1, 1.2) [right, font=\normalsize] {0};
\node (end) at (0.1, 5.2) [right, font=\normalsize] {0};
\node (end) at (0.1, 9.2) [right, font=\normalsize] {0};

%--------G3--------
%----x1----
\node (end) at (2.5, 8.4) [right, font=\normalsize] {0};
\node (end) at (2.5, 7.6) [right, font=\normalsize] {1};
\node (end) at (2.5, 6.8) [right, font=\normalsize] {1};
\node (end) at (3.3, 8.4) [right, font=\normalsize] {1};
\node (end) at (3.3, 7.6) [right, font=\normalsize] {0};
\node (end) at (3.3, 6.8) [right, font=\normalsize] {0};

\node (end) at (2.5, 4.4) [right, font=\normalsize] {0};
\node (end) at (2.5, 3.6) [right, font=\normalsize] {1};
\node (end) at (2.5, 2.8) [right, font=\normalsize] {1};
\node (end) at (3.3, 4.4) [right, font=\normalsize] {1};
\node (end) at (3.3, 3.6) [right, font=\normalsize] {0};
\node (end) at (3.3, 2.8) [right, font=\normalsize] {0};
%----x2----
\node (end) at (5.7, 8.4) [right, font=\normalsize] {0};
\node (end) at (5.7, 7.6) [right, font=\normalsize] {1};
\node (end) at (5.7, 6.8) [right, font=\normalsize] {1};
\node (end) at (6.5, 8.4) [right, font=\normalsize] {1};
\node (end) at (6.5, 7.6) [right, font=\normalsize] {0};
\node (end) at (6.5, 6.8) [right, font=\normalsize] {0};
%----x3----
\node (end) at (8.9, 8.4) [right, font=\normalsize] {0};
\node (end) at (8.9, 7.6) [right, font=\normalsize] {1};
\node (end) at (8.9, 6.8) [right, font=\normalsize] {1};
\node (end) at (9.7, 8.4) [right, font=\normalsize] {1};
\node (end) at (9.7, 7.6) [right, font=\normalsize] {0};
\node (end) at (9.7, 6.8) [right, font=\normalsize] {0};

\node (end) at (8.9, 4.4) [right, font=\normalsize] {0};
\node (end) at (8.9, 3.6) [right, font=\normalsize] {1};
\node (end) at (8.9, 2.8) [right, font=\normalsize] {1};
\node (end) at (9.7, 4.4) [right, font=\normalsize] {1};
\node (end) at (9.7, 3.6) [right, font=\normalsize] {0};
\node (end) at (9.7, 2.8) [right, font=\normalsize] {0};

\node (end) at (12.1, 7.6) [right, font=\normalsize] {0};
\node (end) at (12.1, 6.8) [right, font=\normalsize] {0};
\node (end) at (12.1, 3.6) [right, font=\normalsize] {0};
\node (end) at (12.1, 2.8) [right, font=\normalsize] {0};

\draw[line width=0.6mm] (0, 0.8) -- (12.8, 0.8);
\draw[line width=0.6mm] (0, 2.4) -- (12.8, 2.4);
\draw[line width=0.6mm] (0, 4.8) -- (12.8, 4.8);
\draw[line width=0.6mm] (0, 6.4) -- (12.8, 6.4);

\draw[line width=0.6mm] (0, 8.8) -- (12.8, 8.8);
\draw[line width=0.6mm] (0, 10.4) -- (12.8, 10.4);

\draw[line width=0.6mm] (0, 0.8) -- (0, 2.4);
\draw[line width=0.6mm] (0, 4.8) -- (0, 6.4);
\draw[line width=0.6mm] (0, 8.8) -- (0, 10.4);

\draw[line width=0.6mm] (12.8, 0.8) -- (12.8, 2.4);
\draw[line width=0.6mm] (12.8, 4.8) -- (12.8, 6.4);
\draw[line width=0.6mm] (12.8, 8.8) -- (12.8, 10.4);
\end{tikzpicture}
\vspace{-1.5mm}
\caption{Board after modifying G3}
\label{fig:afterG3}
\vspace{-4.5mm}
\end{figure}
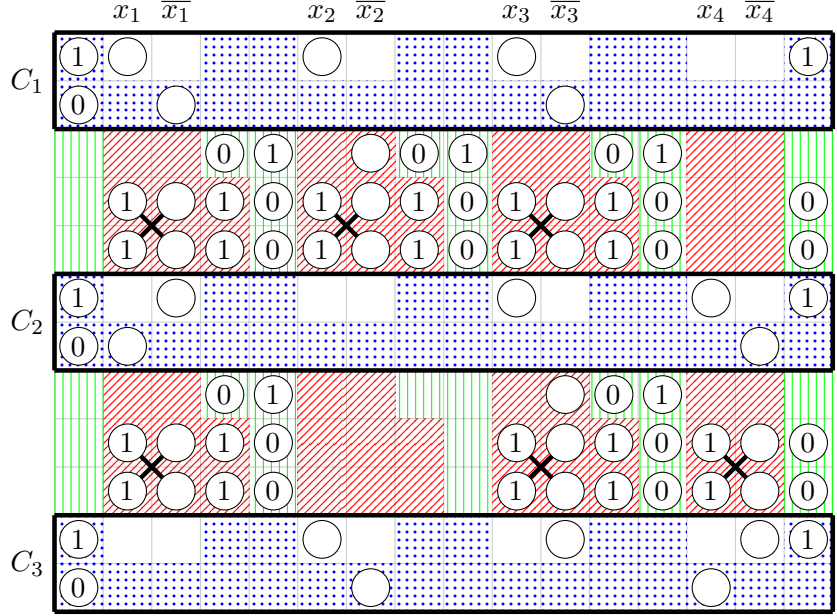

In the board of Figure~\ref{fig:afterG2}, 
the rules for Oredango may be violated due to the areas enclosed by the red, blue, and green frames.
Indeed, each area fails to satisfy the rules as follows:
\begin{enumerate}
\setlength{\leftskip}{5mm}
\item[Red:] Since there are three consecutive black circles in the same column, rule (d) is always violated.
\item[Green:] Since there are three consecutive black circles in the same column, rule (d) is always violated.
\item[Blue:] Depending on the colors of the literal circles, rule (c) can be violated.
\end{enumerate}
We modify the board in gadget G3 to ensure that all the rules are satisfied as follows:
\begin{enumerate}
\setlength{\leftskip}{5mm}
\item[Red:] We insert circles with 0 between the circles with 1.
\item[Green:] We insert circles with 0 between the circles with 1.
\item[Blue:] We insert circles with 0 between the circles with 1.
However, since rule (c) is still not satisfied in the column containing the inserted circle, we further insert a circle with 1 above the inserted circles with~0.
\end{enumerate}
We show the modified board in Figure~\ref{fig:afterG3}.

\vspace{-1mm}

% ---------------- Fig:G3 ----------------

\subsection{Polynomial-time one-to-one reduction}
By the introduction of three gadgets, the desired reduction from 1-in-3SAT to Oredango is obtained.
Clearly, this reduction can be done in polynomial time of the input size of 1-in-3SAT.
From the way of configuration for G1 and G2, Oredango has a solution if and only if there is a solution in 1-in-3SAT.
Moreover, as can be seen from Figure~\ref{fig:afterG3},
given the colors of the literal circles, the colors of the remaining circles are uniquely determined.
As a consequence, through the polynomial-time reduction,
we can establish a one-to-one correspondence between the solution sets of Oredango and 1-in-3SAT. 
Finally, since the length of each skewer and the integer in each circle are at most one, 
as shown in Fig.~\ref{fig:afterG3}, the proof of Theorem~\ref{th:main} is complete.
\vspace{-1.5mm}

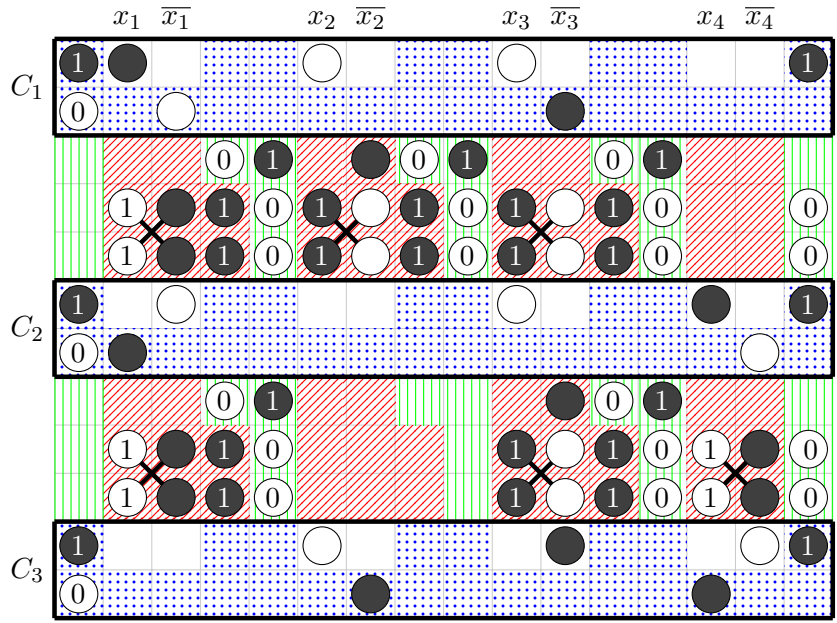
\begin{figure}[H]
\centering
\begin{tikzpicture}[scale=0.8]

%grid
\draw[step=0.8cm, color={rgb:black, 1;white, 4}] (0, 0.8) grid (12.8, 10.4);

\node (end) at (0, 1.6) [left] {$C_{3}$};
\node (end) at (0, 5.6) [left] {$C_{2}$};
\node (end) at (0, 9.6) [left] {$C_{1}$};

\node (end) at (1.2, 10.4) [above] {$x_{1}$};
\node (end) at (2.0, 10.4) [above] {$\overline{x_{1}}$};
\node (end) at (4.4, 10.4) [above] {$x_{2}$};
\node (end) at (5.2, 10.4) [above] {$\overline{x_{2}}$};
\node (end) at (7.6, 10.4) [above] {$x_{3}$};
\node (end) at (8.4, 10.4) [above] {$\overline{x_{3}}$};
\node (end) at (10.8, 10.4) [above] {$x_{4}$};
\node (end) at (11.6, 10.4) [above] {$\overline{x_{4}}$};

%-----------------ガジェットG1-----------------%
\path[pattern=north east lines, pattern color=red] (0.8, 2.4) rectangle + (1.6, 2.4);
\path[pattern=north east lines, pattern color=red] (2.4, 2.4) rectangle + (0.8, 1.6);
\path[pattern=north east lines, pattern color=red] (0.8, 6.4) rectangle + (1.6, 2.4);
\path[pattern=north east lines, pattern color=red] (2.4, 6.4) rectangle + (0.8, 1.6);

\path[pattern=north east lines, pattern color=red] (4.0, 2.4) rectangle + (1.6, 2.4);
\path[pattern=north east lines, pattern color=red] (5.6, 2.4) rectangle + (0.8, 1.6);
\path[pattern=north east lines, pattern color=red] (4.0, 6.4) rectangle + (1.6, 2.4);
\path[pattern=north east lines, pattern color=red] (5.6, 6.4) rectangle + (0.8, 1.6);

\path[pattern=north east lines, pattern color=red] (7.2, 2.4) rectangle + (1.6, 2.4);
\path[pattern=north east lines, pattern color=red] (8.8, 2.4) rectangle + (0.8, 1.6);
\path[pattern=north east lines, pattern color=red] (7.2, 6.4) rectangle + (1.6, 2.4);
\path[pattern=north east lines, pattern color=red] (8.8, 6.4) rectangle + (0.8, 1.6);

\path[pattern=north east lines, pattern color=red] (10.4, 2.4) rectangle + (1.6, 2.4);
\path[pattern=north east lines, pattern color=red] (10.4, 6.4) rectangle + (1.6, 2.4);
%-----------------ガジェットG2-----------------%
\path[pattern=dots, pattern color=blue] (0.0, 1.6) rectangle + (0.8, 0.8);
\path[pattern=dots, pattern color=blue] (0.0, 5.6) rectangle + (0.8, 0.8);
\path[pattern=dots, pattern color=blue] (0.0, 9.6) rectangle + (0.8, 0.8);

\path[pattern=dots, pattern color=blue] (2.4, 1.6) rectangle + (1.6, 0.8);
\path[pattern=dots, pattern color=blue] (2.4, 5.6) rectangle + (1.6, 0.8);
\path[pattern=dots, pattern color=blue] (2.4, 9.6) rectangle + (1.6, 0.8);

\path[pattern=dots, pattern color=blue] (5.6, 1.6) rectangle + (1.6, 0.8);
\path[pattern=dots, pattern color=blue] (5.6, 5.6) rectangle + (1.6, 0.8);
\path[pattern=dots, pattern color=blue] (5.6, 9.6) rectangle + (1.6, 0.8);

\path[pattern=dots, pattern color=blue] (8.8, 1.6) rectangle + (1.6, 0.8);
\path[pattern=dots, pattern color=blue] (8.8, 5.6) rectangle + (1.6, 0.8);
\path[pattern=dots, pattern color=blue] (8.8, 9.6) rectangle + (1.6, 0.8);

\path[pattern=dots, pattern color=blue] (12.0, 1.6) rectangle + (0.8, 0.8);
\path[pattern=dots, pattern color=blue] (12.0, 5.6) rectangle + (0.8, 0.8);
\path[pattern=dots, pattern color=blue] (12.0, 9.6) rectangle + (0.8, 0.8);

\path[pattern=dots, pattern color=blue] (0.0, 0.8) rectangle + (12.8, 0.8);
\path[pattern=dots, pattern color=blue] (0.0, 4.8) rectangle + (12.8, 0.8);
\path[pattern=dots, pattern color=blue] (0.0, 8.8) rectangle + (12.8, 0.8);
%-----------------ガジェットG3-----------------%
\path[pattern=vertical lines, pattern color=green] (0.0, 2.4) rectangle + (0.8, 2.4);
\path[pattern=vertical lines, pattern color=green] (0.0, 6.4) rectangle + (0.8, 2.4);

\path[pattern=vertical lines, pattern color=green] (3.2, 2.4) rectangle + (0.8, 1.6);
\path[pattern=vertical lines, pattern color=green] (2.4, 4.0) rectangle + (1.6, 0.8);
\path[pattern=vertical lines, pattern color=green] (3.2, 6.4) rectangle + (0.8, 1.6);
\path[pattern=vertical lines, pattern color=green] (2.4, 8.0) rectangle + (1.6, 0.8);

\path[pattern=vertical lines, pattern color=green] (6.4, 2.4) rectangle + (0.8, 1.6);
\path[pattern=vertical lines, pattern color=green] (5.6, 4.0) rectangle + (1.6, 0.8);
\path[pattern=vertical lines, pattern color=green] (6.4, 6.4) rectangle + (0.8, 1.6);
\path[pattern=vertical lines, pattern color=green] (5.6, 8.0) rectangle + (1.6, 0.8);

\path[pattern=vertical lines, pattern color=green] (9.6, 2.4) rectangle + (0.8, 1.6);
\path[pattern=vertical lines, pattern color=green] (8.8, 4.0) rectangle + (1.6, 0.8);
\path[pattern=vertical lines, pattern color=green] (9.6, 6.4) rectangle + (0.8, 1.6);
\path[pattern=vertical lines, pattern color=green] (8.8, 8.0) rectangle + (1.6, 0.8);

\path[pattern=vertical lines, pattern color=green] (12.0, 2.4) rectangle + (0.8, 2.4);
\path[pattern=vertical lines, pattern color=green] (12.0, 6.4) rectangle + (0.8, 2.4);
%-----------串の描画-----------
\draw[line width=0.6mm] (1.2, 6.8) -- (2.0, 7.6);
\draw[line width=0.6mm] (1.2, 7.6) -- (2.0, 6.8);
\draw[line width=0.6mm] (1.2, 2.8) -- (2.0, 3.6);
\draw[line width=0.6mm] (1.2, 3.6) -- (2.0, 2.8);

\draw[line width=0.6mm] (4.4, 6.8) -- (5.2, 7.6);
\draw[line width=0.6mm] (4.4, 7.6) -- (5.2, 6.8);

\draw[line width=0.6mm] (7.6, 6.8) -- (8.4, 7.6);
\draw[line width=0.6mm] (7.6, 7.6) -- (8.4, 6.8);
\draw[line width=0.6mm] (7.6, 2.8) -- (8.4, 3.6);
\draw[line width=0.6mm] (7.6, 3.6) -- (8.4, 2.8);

\draw[line width=0.6mm] (10.8, 2.8) -- (11.6, 3.6);
\draw[line width=0.6mm] (10.8, 3.6) -- (11.6, 2.8);
%-----------だんごの描画-----------
%C_1
\node[draw, circle, minimum size=0.5cm, fill=blk] at (1.2, 10.0) {};
\node[draw, circle, minimum size=0.5cm, fill=wht] at (2.0, 9.2) {};
\node[draw, circle, minimum size=0.5cm, fill=wht] at (4.4, 10.0) {};
\node[draw, circle, minimum size=0.5cm, fill=blk] at (5.2, 8.4) {};
\node[draw, circle, minimum size=0.5cm, fill=wht] at (0.4, 9.2) {};
\node[draw, circle, minimum size=0.5cm, fill=wht] at (7.6, 10.0) {};
\node[draw, circle, minimum size=0.5cm, fill=blk] at (8.4, 9.2) {};
%C2
\node[draw, circle, minimum size=0.5cm, fill=blk] at (1.2, 5.2) {};
\node[draw, circle, minimum size=0.5cm, fill=wht] at (2.0, 6.0) {};
\node[draw, circle, minimum size=0.5cm, fill=wht] at (7.6, 6.0) {};
\node[draw, circle, minimum size=0.5cm, fill=blk] at (8.4, 4.4) {};
\node[draw, circle, minimum size=0.5cm, fill=wht] at (0.4, 5.2) {};
\node[draw, circle, minimum size=0.5cm, fill=blk] at (10.8, 6.0) {};
\node[draw, circle, minimum size=0.5cm, fill=wht] at (11.6, 5.2) {};
%C3
\node[draw, circle, minimum size=0.5cm, fill=wht] at (4.4, 2.0) {};
\node[draw, circle, minimum size=0.5cm, fill=blk] at (5.2, 1.2) {};
\node[draw, circle, minimum size=0.5cm, fill=blk] at (8.4, 2.0) {};
\node[draw, circle, minimum size=0.5cm, fill=wht] at (0.4, 1.2) {};
\node[draw, circle, minimum size=0.5cm, fill=blk] at (10.8, 1.2) {};
\node[draw, circle, minimum size=0.5cm, fill=wht] at (11.6, 2.0) {};
%--------G1--------
%----x1----
\node[draw, circle, minimum size=0.5cm, fill=wht] at (1.2, 7.6) {};
\node[draw, circle, minimum size=0.5cm, fill=blk] at (2.0, 7.6) {};
\node[draw, circle, minimum size=0.5cm, fill=blk] at (2.8, 7.6) {};
\node[draw, circle, minimum size=0.5cm, fill=wht] at (1.2, 6.8) {};
\node[draw, circle, minimum size=0.5cm, fill=blk] at (2.0, 6.8) {};
\node[draw, circle, minimum size=0.5cm, fill=blk] at (2.8, 6.8) {};
\node[draw, circle, minimum size=0.5cm, fill=wht] at (1.2, 3.6) {};
\node[draw, circle, minimum size=0.5cm, fill=blk] at (2.0, 3.6) {};
\node[draw, circle, minimum size=0.5cm, fill=blk] at (2.8, 3.6) {};
\node[draw, circle, minimum size=0.5cm, fill=wht] at (1.2, 2.8) {};
\node[draw, circle, minimum size=0.5cm, fill=blk] at (2.0, 2.8) {};
\node[draw, circle, minimum size=0.5cm, fill=blk] at (2.8, 2.8) {};
%----x2----
\node[draw, circle, minimum size=0.5cm, fill=blk] at (4.4, 7.6) {};
\node[draw, circle, minimum size=0.5cm, fill=wht] at (5.2, 7.6) {};
\node[draw, circle, minimum size=0.5cm, fill=blk] at (6.0, 7.6) {};
\node[draw, circle, minimum size=0.5cm, fill=blk] at (4.4, 6.8) {};
\node[draw, circle, minimum size=0.5cm, fill=wht] at (5.2, 6.8) {};
\node[draw, circle, minimum size=0.5cm, fill=blk] at (6.0, 6.8) {};
%----x3----
\node[draw, circle, minimum size=0.5cm, fill=blk] at (7.6, 7.6) {};
\node[draw, circle, minimum size=0.5cm, fill=wht] at (8.4, 7.6) {};
\node[draw, circle, minimum size=0.5cm, fill=blk] at (9.2, 7.6) {};
\node[draw, circle, minimum size=0.5cm, fill=blk] at (7.6, 6.8) {};
\node[draw, circle, minimum size=0.5cm, fill=wht] at (8.4, 6.8) {};
\node[draw, circle, minimum size=0.5cm, fill=blk] at (9.2, 6.8) {};
\node[draw, circle, minimum size=0.5cm, fill=blk] at (7.6, 3.6) {};
\node[draw, circle, minimum size=0.5cm, fill=wht] at (8.4, 3.6) {};
\node[draw, circle, minimum size=0.5cm, fill=blk] at (9.2, 3.6) {};
\node[draw, circle, minimum size=0.5cm, fill=blk] at (7.6, 2.8) {};
\node[draw, circle, minimum size=0.5cm, fill=wht] at (8.4, 2.8) {};
\node[draw, circle, minimum size=0.5cm, fill=blk] at (9.2, 2.8) {};
%----x4----
\node[draw, circle, minimum size=0.5cm, fill=wht] at (10.8, 3.6) {};
\node[draw, circle, minimum size=0.5cm, fill=blk] at (11.6, 3.6) {};
\node[draw, circle, minimum size=0.5cm, fill=wht] at (10.8, 2.8) {};
\node[draw, circle, minimum size=0.5cm, fill=blk] at (11.6, 2.8) {};
%--------G2--------
\node[draw, circle, minimum size=0.5cm, fill=blk] at (0.4, 10.0) {};
\node[draw, circle, minimum size=0.5cm, fill=blk] at (12.4, 10.0) {};
\node[draw, circle, minimum size=0.5cm, fill=blk] at (0.4, 6.0) {};
\node[draw, circle, minimum size=0.5cm, fill=blk] at (12.4, 6.0) {};
\node[draw, circle, minimum size=0.5cm, fill=blk] at (0.4, 2.0) {};
\node[draw, circle, minimum size=0.5cm, fill=blk] at (12.4, 2.0) {};
%--------G3--------
%左端

%----x1----
\node[draw, circle, minimum size=0.5cm, fill=wht] at (2.8, 8.4) {};
\node[draw, circle, minimum size=0.5cm, fill=blk] at (3.6, 8.4) {};
\node[draw, circle, minimum size=0.5cm, fill=wht] at (3.6, 7.6) {};
\node[draw, circle, minimum size=0.5cm, fill=wht] at (3.6, 6.8) {};

\node[draw, circle, minimum size=0.5cm, fill=wht] at (2.8, 4.4) {};
\node[draw, circle, minimum size=0.5cm, fill=blk] at (3.6, 4.4) {};
\node[draw, circle, minimum size=0.5cm, fill=wht] at (3.6, 3.6) {};
\node[draw, circle, minimum size=0.5cm, fill=wht] at (3.6, 2.8) {};
%----x2----
\node[draw, circle, minimum size=0.5cm, fill=wht] at (6.0, 8.4) {};
\node[draw, circle, minimum size=0.5cm, fill=blk] at (6.8, 8.4) {};
\node[draw, circle, minimum size=0.5cm, fill=wht] at (6.8, 7.6) {};
\node[draw, circle, minimum size=0.5cm, fill=wht] at (6.8, 6.8) {};
%----x3----
\node[draw, circle, minimum size=0.5cm, fill=wht] at (9.2, 8.4) {};
\node[draw, circle, minimum size=0.5cm, fill=blk] at (10.0, 8.4) {};
\node[draw, circle, minimum size=0.5cm, fill=wht] at (10.0, 7.6) {};
\node[draw, circle, minimum size=0.5cm, fill=wht] at (10.0, 6.8) {};

\node[draw, circle, minimum size=0.5cm, fill=wht] at (9.2, 4.4) {};
\node[draw, circle, minimum size=0.5cm, fill=blk] at (10.0, 4.4) {};
\node[draw, circle, minimum size=0.5cm, fill=wht] at (10.0, 3.6) {};
\node[draw, circle, minimum size=0.5cm, fill=wht] at (10.0, 2.8) {};

\node[draw, circle, minimum size=0.5cm, fill=wht] at (12.4, 7.6) {};
\node[draw, circle, minimum size=0.5cm, fill=wht] at (12.4, 6.8) {};
\node[draw, circle, minimum size=0.5cm, fill=wht] at (12.4, 3.6) {};
\node[draw, circle, minimum size=0.5cm, fill=wht] at (12.4, 2.8) {};
%--------G1--------
%----x1----
\node (end) at (0.9, 7.6) [right, font=\normalsize] {1};
\node (end) at (0.9, 6.8) [right, font=\normalsize] {1};
\node (end) at (0.9, 3.6) [right, font=\normalsize] {1};
\node (end) at (0.9, 2.8) [right, font=\normalsize] {1};
%----x2----
\node (end) at (4.1, 7.6) [right, font=\normalsize, wht] {1};
\node (end) at (4.1, 6.8) [right, font=\normalsize, wht] {1};
%----x3----
\node (end) at (7.3, 7.6) [right, font=\normalsize, wht] {1};
\node (end) at (7.3, 6.8) [right, font=\normalsize, wht] {1};
\node (end) at (7.3, 3.6) [right, font=\normalsize, wht] {1};
\node (end) at (7.3, 2.8) [right, font=\normalsize, wht] {1};
%----x4----
\node (end) at (10.5, 3.6) [right, font=\normalsize] {1};
\node (end) at (10.5, 2.8) [right, font=\normalsize] {1};
%--------G2--------
\node (end) at (0.1, 2.0) [right, font=\normalsize, wht] {1};
\node (end) at (12.1, 2.0) [right, font=\normalsize, wht] {1};
\node (end) at (0.1, 6.0) [right, font=\normalsize, wht] {1};
\node (end) at (12.1, 6.0) [right, font=\normalsize, wht] {1};
\node (end) at (0.1, 10.0) [right, font=\normalsize, wht] {1};
\node (end) at (12.1, 10.0) [right, font=\normalsize, wht] {1};

\node (end) at (0.1, 1.2) [right, font=\normalsize] {0};
\node (end) at (0.1, 5.2) [right, font=\normalsize] {0};
\node (end) at (0.1, 9.2) [right, font=\normalsize] {0};

%--------G3--------
%----x1----
\node (end) at (2.5, 8.4) [right, font=\normalsize] {0};
\node (end) at (2.5, 7.6) [right, font=\normalsize, wht] {1};
\node (end) at (2.5, 6.8) [right, font=\normalsize, wht] {1};
\node (end) at (3.3, 8.4) [right, font=\normalsize, wht] {1};
\node (end) at (3.3, 7.6) [right, font=\normalsize] {0};
\node (end) at (3.3, 6.8) [right, font=\normalsize] {0};

\node (end) at (2.5, 4.4) [right, font=\normalsize] {0};
\node (end) at (2.5, 3.6) [right, font=\normalsize, wht] {1};
\node (end) at (2.5, 2.8) [right, font=\normalsize, wht] {1};
\node (end) at (3.3, 4.4) [right, font=\normalsize, wht] {1};
\node (end) at (3.3, 3.6) [right, font=\normalsize] {0};
\node (end) at (3.3, 2.8) [right, font=\normalsize] {0};
%----x2----
\node (end) at (5.7, 8.4) [right, font=\normalsize] {0};
\node (end) at (5.7, 7.6) [right, font=\normalsize, wht] {1};
\node (end) at (5.7, 6.8) [right, font=\normalsize, wht] {1};
\node (end) at (6.5, 8.4) [right, font=\normalsize, wht] {1};
\node (end) at (6.5, 7.6) [right, font=\normalsize] {0};
\node (end) at (6.5, 6.8) [right, font=\normalsize] {0};
%----x3----
\node (end) at (8.9, 8.4) [right, font=\normalsize] {0};
\node (end) at (8.9, 7.6) [right, font=\normalsize, wht] {1};
\node (end) at (8.9, 6.8) [right, font=\normalsize, wht] {1};
\node (end) at (9.7, 8.4) [right, font=\normalsize, wht] {1};
\node (end) at (9.7, 7.6) [right, font=\normalsize] {0};
\node (end) at (9.7, 6.8) [right, font=\normalsize] {0};

\node (end) at (8.9, 4.4) [right, font=\normalsize] {0};
\node (end) at (8.9, 3.6) [right, font=\normalsize, wht] {1};
\node (end) at (8.9, 2.8) [right, font=\normalsize, wht] {1};
\node (end) at (9.7, 4.4) [right, font=\normalsize, wht] {1};
\node (end) at (9.7, 3.6) [right, font=\normalsize] {0};
\node (end) at (9.7, 2.8) [right, font=\normalsize] {0};

\node (end) at (12.15, 7.6) [right, font=\normalsize] {0};
\node (end) at (12.15, 6.8) [right, font=\normalsize] {0};
\node (end) at (12.15, 3.6) [right, font=\normalsize] {0};
\node (end) at (12.15, 2.8) [right, font=\normalsize] {0};

\draw[line width=0.6mm] (0, 0.8) -- (12.8, 0.8);
\draw[line width=0.6mm] (0, 2.4) -- (12.8, 2.4);
\draw[line width=0.6mm] (0, 4.8) -- (12.8, 4.8);
\draw[line width=0.6mm] (0, 6.4) -- (12.8, 6.4);

\draw[line width=0.6mm] (0, 8.8) -- (12.8, 8.8);
\draw[line width=0.6mm] (0, 10.4) -- (12.8, 10.4);

\draw[line width=0.6mm] (0, 0.8) -- (0, 2.4);
\draw[line width=0.6mm] (0, 4.8) -- (0, 6.4);
\draw[line width=0.6mm] (0, 8.8) -- (0, 10.4);

\draw[line width=0.6mm] (12.8, 0.8) -- (12.8, 2.4);
\draw[line width=0.6mm] (12.8, 4.8) -- (12.8, 6.4);
\draw[line width=0.6mm] (12.8, 8.8) -- (12.8, 10.4);
\end{tikzpicture}
\vspace{-2mm}
\caption{Answer for the input in Figure~\ref{fig:afterG3}}
\label{fig:AnsG3}
\vspace{-4mm}
\end{figure}

\begin{example}
\label{ex:1in3sattoRedango}
A solution for the 1-in-3SAT input of Example~\ref{ex:1in3sat} is
\begin{align}
\label{eq:ans}
    x_1=1, \ x_2=0, \ x_3=0, \ x_4=1.
\end{align}
Figure~\ref{fig:AnsG3} shows the unique solution for the input in Figure~\ref{fig:afterG3} 
and the colors of literal circles corresponding to \eqref{eq:ans}.
This solution satisfies rules {\rm (a)-(d)}. All the literal circles 
corresponding to $x_1$ and $x_4$ {\rm(}$\overline{x_1}$ and $\overline{x_4}${\rm)} are black (white),
and all the literal circles
corresponding to $x_2$ and $x_3$ {\rm(}$\overline{x_2}$ and $\overline{x_3}${\rm)} are white (black).
\end{example}

\vspace{-4mm}

% ---------------- Fig:AnsG3 ----------------

\section{0-1 Integer Programming Formulation}\label{sec:IP}
In this section, we formulate Oredango as a 0-1 integer programming problem.
We begin by introducing a 0-1 variable $x_{i,j}$ for each cell at $(i,j)$
to express that the circle at $(i,j)$ is black or white, that is,
\begin{align*}
x_{i,j} &= 1 \quad \mbox{means that the circle at $(i,j)$ is black},\\
x_{i,j} &= 0 \quad \mbox{means that the circle at $(i,j)$ is white}.
\end{align*}
Recall the rules (a)-(d) defined in Section \ref{sec:intro}.
Using these variables, we express rules (a)-(d) as four linear constraints.

%----------------- Constraints for skewers -----------------%
\medskip
\noindent \textbf{Constraints for skewers (rules (a) and (b))}

Let $S_1, S_2, \cdots, S_k$ be the $k~(\ge 0)$ skewers given as input.
We denote by $a_r$ the integer written on the $S_r$ skewer,
and also denote by $s_r$ the number of the circles connected through the skewer~$S_r$.
Moreover, we denote the coordinates of the circles connected through the skewer~$S_r$
by
$(i^r_1, j^r_1), (i^r_2, j^r_2), \cdots,(i^r_{s_r}, j^r_{s_r})$.
Here, we suppose 
for any $s=1,2,\ldots,s_r-1$,
the two circles indicated by 
$(i^r_s,j^r_s)$ and $(i^r_{s+1},j^r_{s+1})$ 
are adjacent to each other. 

Rule (a) can be expressed by the following linear constraint inequality straightforwardly:
\begin{align}
\label{eq:rule_a}
\sum\limits_{1 \le t \le s_r}x_{i^r_{t}, j^r_{t}}=a_{r} \quad (r=1, \ldots , k).
\end{align}
Rule (b) can be expressed as
\begin{align}
\label{eq:rule_b}
1\leq x_{i^{r}_{t}, j^{r}_{t}} &+ x_{i^{r}_{t+1}, j^{r}_{t+1}} + x_{i^{r}_{t+2}, j^{r}_{t+2}} \leq 2 \nonumber\\
& (r=1, \ldots, k, t=1, \ldots, s_r-2),
\end{align}
where the middle expression stands 
for the number of three consecutive black circles from the $t$-th to $(t+2)$-th.
This inequality straightforwardly implies that, in any three consecutive circles,
there are one or two black circles, and there are no three consecutive white (black) circles. 

%----------------- Constraints for rows -----------------%
\medskip
\noindent \textbf{Constraint for rows (rule (c))}

For $i \in \{1, \ldots, m\}$, 
to indicate any three consecutive circles on the $i$-th row,
we define a family of sets $\mathcal{C}^i = \{C^i_1, C^i_2, \ldots \}$
where each $C^i_j$ consists of the coordinates of the three circles arranged consecutively on the $i$-th row.
For instance,
as for the $1$st and 3rd rows in the left figure of Fig~1, 
we have 
$\mathcal{C}^1 = \{C^1_1, C^1_2\}$ 
with $C^1_1=\{(1,1),(1,2),(1,3)\}$ and 
$C^1_2=\{(1,2),(1,3),(1,4)\}$, and also have
$\mathcal{C}^3 = \{C^3_1\}$ with $C^3_1=\{(3,1),(3,2),(3,4)\}$.
By definition, $|C^i_1| = |C^i_2| = \cdots = 3$ holds,
and if there are less than three circles in the $i$-th row, then $\mathcal{C}^i = \emptyset$ holds.
Under this setting, rule (c) is represented as
\begin{align}
\label{eq:rule_c}
\hspace{-4mm}
1 \leq \sum\limits_{(i, j)\in C^i_{s}}x_{i, j} \leq 2 \ 
(i=1, \ldots, m, \ s = 1, \ldots, |\mathcal{C}^i|).
\end{align}

%----------------- Constraints for columns -----------------%
\medskip
\noindent \textbf{Constraint for columns (rule (d))}

For columns, in the same manner as in rule (c) regarding rows,
we define a family of sets $\mathcal{D}^j = \{D^j_1, D^j_2, \ldots \}$ for $j \in \{1, \ldots, n\}$.
Rule (d) is represented as
\begin{align}
\label{eq:rule_d}
\hspace{-4mm}
1 \leq \sum\limits_{(i, j)\in D^j_{s}}x_{i, j} \leq 2 \ 
(j=1, \ldots, n, \ t = 1, \ldots, |\mathcal{D}^j|).
\end{align}
Since solving Oredango is simply finding a solution which satisfies rules (a)--(d), 
it is equivalent to solving a 0-1 integer program with constraints
\eqref{eq:rule_a}--\eqref{eq:rule_d}, along with an arbitrary linear objective function. 

\section{Experimental Results}\label{sec:result}

In this section, we show the results of solving 36 puzzles
(11 problems from the puzzle magazine Nikoli \cite{Nikoli}, 25 problems from PuzzleSquareJP \cite{PS} (PS for short))
using the formulation as a 0-1 integer programming problem 
based on inequalities \eqref{eq:rule_a} to \eqref{eq:rule_d}.
We set the summation of all the 0-1 variables $x_{i,j}$ to the objective function.
For the solution of integer programming problem,
we use Gurobi Version 10.0.3 called from C++.
Our experiments were conducted on an Intel Core i5-11400 with 2.60GHz and 32GB RAM.

We show the results in Table~\ref{TB:IP}.
``Size'' is the width and height of the board, 
and ``\# of circles'' is the number of circles on the initial board.
In ``Reference'',
if the instance is from Nikoli, we show the magazine volume and problem number,
and if the instance is from PS, 
we show the pid associated with the end of the URL of PS.
For example, if the reference is ``PS pid:113543'',
the instance can be found at 
``https://puzsq.logicpuzzle.app/puzzle/113543.''
Each of the instances was solved in less than one second in our experiments;
thus, we confirm that the approach using 0-1 integer programming formulation 
and Gurobi can solve the problem of Nikoli and PS very efficiently. 

\begin{table}[H]
 \centering 
 \caption{Experimental results for the instances at Nikoli and PS}
 \scriptsize
  \begin{tabular}{|c|c|c|c|l|}
   \hline ${\rule{0pt}{2.7mm}}$
   \textbf{ID} & \textbf{Size} & \textbf{\# of circles} &\textbf{Time (sec)}& \multicolumn{1}{c|}{\textbf{Reference}}\\
   \hline 
   1 & 3 $\times$ 3 & 8 &0.196&PS pid:113543\\
   2 & 3 $\times$ 3 & 8 &0.189&PS pid:113549\\
   3 & 3 $\times$ 3 & 8 &0.224&Nikoli vol.184 Q1\\
   4 & 3 $\times$ 3 & 9 &0.210&PS pid:113531\\
   5 & 3 $\times$ 3 & 9 &0.245&PS pid:113547\\
   6 & 4 $\times$ 3 & 12 &0.218&PS pid:113791\\
   7 & 4 $\times$ 3 & 12 &0.284&PS pid:113793\\
   8 & 4 $\times$ 3 & 12 &0.257&PS pid:113980\\
   9 & 4 $\times$ 4 & 11 &0.189&PS pid:112956\\
   10 & 4 $\times$ 4 & 12 &0.221&Nikoli vol.184 Q2\\
   11 & 4 $\times$ 4 & 13 &0.201&PS pid:113549\\
   12 & 4 $\times$ 4 & 14 &0.247&PS pid:112814\\
   13 & 4 $\times$ 4 & 14 &0.228&Nikoli vol.184 Q3\\
   14 & 4 $\times$ 4 & 14 &0.321&Nikoli vol.185 Q1\\
   15 & 4 $\times$ 4 & 15 &0.207&PS pid:113609\\
   16 & 4 $\times$ 4 & 16 &0.197&PS pid:113751\\
   17 & 5 $\times$ 4 & 18 &0.195&PS pid:113980\\
   18 & 5 $\times$ 4 & 18 &0.296&PS pid:121914\\
   19 & 5 $\times$ 4 & 19 &0.208&PS pid:113231\\
   20 & 5 $\times$ 4 & 20 &0.185&PS pid:121857\\
   21 & 6 $\times$ 4 & 22 &0.175&PS pid:121882\\
   22 & 5 $\times$ 5 & 16 &0.266&PS pid:112984\\
   23 & 5 $\times$ 5 & 16 &0.226&Nikoli vol.185 Q2\\
   24 & 5 $\times$ 5 & 19 &0.249&PS pid:112782\\
   25 & 5 $\times$ 5 & 21 &0.221&Nikoli vol.184 Q4\\
   26 & 5 $\times$ 5 & 24 &0.265&PS pid:114122\\
   27 & 5 $\times$ 5 & 25 &0.366&PS pid:115053\\
   28 & 5 $\times$ 5 & 25 &0.270&PS pid:115069\\
   29 & 6 $\times$ 6 & 21 &0.237&PS pid:114395\\
   30 & 6 $\times$ 6 & 25 &0.262&PS pid:113006\\
   31 & 6 $\times$ 6 & 32 &0.268&Nikoli vol.185 Q7\\
   32 & 6 $\times$ 6 & 35 &0.267&PS pid:114158\\
   33 & 7 $\times$ 7 & 29 &0.201&Nikoli vol.185 Q3\\
   34 & 7 $\times$ 7 & 41 &0.221&Nikoli vol.185 Q4\\
   35 & 8 $\times$ 8 & 41 &0.246&Nikoli vol.185 Q6\\
   36 & 10 $\times$ 10 & 76 &0.433&Nikoli vol.185 Q5\\
   \hline
  \end{tabular}
 \label{TB:IP}
\end{table}

\section*{Acknowledgement}
This work was supported by JSPS KAKENHI Grant Number JP23K16842.

\end{document}